\documentclass[11pt, a4paper]{article}
\pdfoutput=1

\usepackage[utf8]{inputenc}
\usepackage{jcappub}

\usepackage{amssymb,amsmath,latexsym,mathrsfs}
\usepackage{xcolor}
\usepackage{bm}
\usepackage{url}
\usepackage{epsfig,graphicx,verbatim,xspace,multirow,mathtools}
\usepackage{array}
\usepackage{ulem}
\usepackage{mathtools}
\newcommand{\ph}{\;}

\newcommand{\eq}{\mathrm{eq}}
\newcommand{\D}{\text{d}}

\newcommand{\q}{\left(}
\newcommand{\ie}{\emph{i.e.}}
\newcommand{\eg}{\emph{e.g.}}

\newcommand{\w}{\right)}

\newcommand{\mpl}{M_\mathrm{Pl}}
\newcommand{\mref}{m_{\rm ref}}
\newcommand{\TFI}{T_{\rm FI}}

\newcommand{\TSW}{T_{\rm SW}}

\newcommand{\Tprod}{T_{\rm prod}}
\newcommand{\aprod}{a_{\rm prod}}
\newcommand{\qav}{\langle q\rangle}
\newcommand{\qnav}{\langle q^n\rangle}
\newcommand{\gprod}{g_{*S}(t_{\rm prod})}
\newcommand{\gtod}{g_{*S}(t_0)}
\newcommand{\xFI}{x_{\rm FI}}

\newcommand{\TEW}{T_{\rm EW}}
\newcommand{\TFO}{T_{\rm FO}}
\newcommand{\gs}{g_{*S}}
\newcommand{\Neff}{N_{\rm eff}}
\newcommand{\dAw}{\delta A_{\rm WDM}}
\newcommand{\class}{\textsc{class}}
\newcommand{\vrel}{v_{\rm rel}}

\title{Lyman-$\alpha$ constraints on freeze-in and superWIMPs}

\author[a]{Quentin Decant,}
\emailAdd{quentin.decant@ulb.be}
\author[b,c]{Jan Heisig,}
\emailAdd{heisig@physik.rwth-aachen.de}
\author[d]{Deanna C. Hooper,}
\emailAdd{deanna.hooper@helsinki.fi}
\author[a,e]{\\ and Laura Lopez-Honorez}
\emailAdd{llopezho@ulb.be}

\affiliation[a]{Service de Physique Th\'eorique, CP225, Universit\'e Libre de Bruxelles, Bld du Triomphe, B-1050 Brussels, Belgium}

\affiliation[b]{Institute for Theoretical Particle Physics and Cosmology, RWTH Aachen University, Sommerfeldstr. 16, D-52056 Aachen, Germany}
\affiliation[c]{Centre for Cosmology, Particle Physics and Phenomenology (CP3), Universit\'e catholique de Louvain, B-1348 Louvain-la-Neuve, Belgium}

\affiliation[d]{Department of Physics and Helsinki Institute of Physics, PL 64, FI-00014 University of Helsinki, Finland}

\affiliation[e]{Theoretische Natuurkunde, Vrije Universiteit Brussel and The International Solvay Institutes,
Pleinlaan 2, B-1050 Brussels, Belgium}

\abstract{ 
Dark matter (DM) from freeze-in or superWIMP production is well known to imprint non-cold DM signatures on cosmological observables. We derive constraints from Lyman-$\alpha$ forest observations for both cases, basing ourselves on a reinterpretation of the existing Lyman-$\alpha$ limits on thermal warm DM\@. We exclude  DM masses below  15 keV for freeze-in, in good agreement with previous literature, and provide a generic lower mass bound for superWIMPs that depends on the mother particle decay width. Special emphasis is placed on the mixed scenario, where contributions from both freeze-in and superWIMP are similarly important. In this case, the imprint on cosmological observables can deviate significantly from thermal warm DM\@. Furthermore, we provide a modified version of the Boltzmann code \class, analytic expressions for the DM distributions, and fits to the DM transfer functions that account for both mechanisms of production. Moreover, we also derive generic constraints from $\Delta N_\mathrm{eff}$ measurements and show that they cannot compete with those arising from \mbox{Lyman-$\alpha$} observations. For illustration, we apply the above generic limits to a coloured $t$-channel mediator DM model, in which case contributions from both freeze-in through scatterings and decays, as well as superWIMP production can be important. We map out the entire cosmologically viable parameter space, cornered by bounds from Lyman-$\alpha$ observations, the LHC, and Big Bang Nucleosynthesis.
  }

\begin{document}

\hfill{\small ULB-TH/21-20}

\hfill{\small TTK-21-46}

\hfill{\small HIP-2021-38/TH}

\maketitle

\section{Introduction}
\label{sec:intro}
Cosmological observations imply that around 80\% of the total matter
content in our universe is made up of dark matter
(DM)~\cite{Planck:2018vyg}.  The gravitational impact of DM on the
dynamics of visible matter has been measured on a large range of
astrophysical and cosmological scales.  Nonetheless, despite
substantial effort, searches in colliders~\cite{Kahlhoefer:2017dnp},
direct~\cite{Undagoitia:2015gya}, and indirect~\cite{Gaskins:2016cha}
experiments have so far not yielded any clear hints of interactions
other than gravitational between the DM and the standard model
particles.

While the aforementioned search strategies depend on the existence
(and sufficient strength) of such an interaction, here we focus on a
complementary path to constrain particle physics models of DM, by
considering the DM imprint on the formation of cosmological structures
and their potential contribution to the effective number of
  neutrinos, $\Delta \Neff$. This is of particular relevance for very
weakly interacting DM, potentially out-of-reach of other search
strategies.

An especially relevant probe in this direction is the Lyman-$\alpha$
forest, which provides a measurement of the positions of hydrogen
clouds along the line-of-sight through the absorption lines of distant
quasars~\cite{Viel:2005qj, Viel:2013fqw,
  Palanque-Delabrouille:2019iyz, Garzilli:2019qki}.  Accordingly,
Lyman-$\alpha$ forest observations probe structure on intermediate
to small scales at redshifts around $2\!\sim
\!6$~\cite{Ikeuchi:1986qq, 10.1093/mnras/218.1.25P}. These small-scale
structures can be washed out by DM free-streaming, which is caused by
significant deviations in the DM momentum distribution compared to the
standard cold dark matter (CDM) scenario.  Various groups have
analysed data of the Lyman-$\alpha$ flux power
spectrum~\cite{Viel:2013fqw,Irsic:2017ixq,Palanque-Delabrouille:2019iyz,Garzilli:2019qki}
and provided results for canonical warm dark matter (WDM),
\ie~thermalised DM that freezes out relativistically in the early
universe. In this scenario, masses below 5.3
keV~\cite{Palanque-Delabrouille:2019iyz} could be excluded under
reasonable assumptions, see however~\cite{Garzilli:2019qki} for a
critical discussion of these assumptions, where the bound is then
reduced to 1.9 keV. 

Here we consider non-thermalised DM, \ie~a DM candidate that is so weakly coupled to the standard model that it never reaches thermal equilibrium with the primordial plasma of standard model particles. Such candidates 
are commonly referred to as \emph{feebly interacting massive particles} (FIMPs).
In these scenarios we can, therefore, no longer rely on the standard freeze-out mechanism to produce the correct relic abundance of DM\@. However, despite its feeble interaction, DM may still be produced to a sufficient amount  
by scatterings or decays of other (thermalised) particles. 
There are mainly two such production mechanisms that have been considered in the literature. 
\emph{(i)} Freeze-in (FI)~\cite{Bolz:2000fu,Pradler:2006hh,McDonald:2001vt,Covi:2002vw,Asaka:2005cn,Frere:2006hp,Hall:2009bx} is the non-efficient production of DM from decays or scatterings of particles in the thermal bath, where \emph{non-efficient} refers to the fact that the respective production rate is small compared to the Hubble expansion rate. 
\emph{(ii)} The superWIMP (SW) mechanism~\cite{Covi:1999ty,Feng:2003uy} is the late decay of a frozen-out mother particle into DM\@. 
While both contributions may arise from the very same decay process, typically they take place at very different times. Hence, their characteristic momentum distribution -- relevant for their imprint on cosmological structures -- can be very different. 

As the scales considered by Lyman-$\alpha$ data lie in the non-linear
regime, normally assessing the impact of a certain DM model on the
Lyman-$\alpha$ forest requires computationally expensive hydrodynamic
simulations.  However, on the basis of only the linear matter power
spectrum -- which we obtain from a modified version of the Boltzmann
code \textsc{class}~\cite{Blas:2011rf, Lesgourgues:2011rh} -- we can,
to good approximation, use the results obtained for WDM to estimate
Lyman-$\alpha$ constraints for the model considered here.  To do so,
we employ three different strategies, with varying degrees of
sophistication and uncertainty.  First, following the approach
of~\cite{Bae:2017dpt}, we consider the velocity dispersion as the
characteristic measure of the free-streaming of DM\@.  Second, we use
an analytical fit to the transfer function, which relates the linear
matter power spectrum of a model to a CDM one, and constrain the
fitting parameters, as was done in~\cite{Bode:2000gq,Viel:2005qj}.
Finally, we make use of the area
criterion~\cite{Murgia:2017lwo,Schneider:2016uqi}, which considers the
integral over the one-dimensional linear power spectrum as a
characteristic quantity constrained by Lyman-$\alpha$ data. Although
all three methods will allow us to derive limits on the pure FI or SW
case, only the latter enables the analysis of the mixed
scenario. In our analyses, we also study the conditions under
which the FIMPs considered here could give rise to significant
contributions to $\Delta \Neff$, reaching the conclusion that this is not expected to provide any more stringent constraints on
the FI or SW scenarios.

Having derived general bounds for these models, we then consider a benchmark scenario with a top-philic simplified $t$-channel mediator model introducing a coloured scalar top-partner and a singled Majorana DM candidate, both odd under a discrete $Z_2$-symmetry that stabilises DM\@. We thereby extend the work of~\cite{Garny:2018ali}, where Lyman-$\alpha$ constraints on the model were estimated by simple considerations of the free-streaming length. Furthermore, following~\cite{Harz:2018csl}, we take into account important bound state formation effects in the freeze-out process of the mediator, which are particularly relevant for the computation of the Lyman-$\alpha$ constraints towards high mediator masses. 

This paper is organised as follows. We begin in Sec.~\ref{sec:FIbasics} by discussing the different production mechanisms for FIMPs, as well as the corresponding Boltzmann equations. In Sec.~\ref{sec:cosmo}, we focus on the cosmological implications of FIMP DM, reviewing the observables that will constrain these models. We then focus on a specific realisation of our set-up, top-philic FIMPs, in Sec.~\ref{sec:topphilic}, before concluding in Sec.~\ref{sec:concl}. Finally, in App.~\ref{sec:more} we go into more detail about SW production, in App,~\ref{sec:somBSF} we provide all relevant expressions for Sommerfeld enhancement and bound state formation, and in App.~\ref{sec:fit_fluid} we discuss the various approximations and consideration made to extract the Lyman-$\alpha$ bounds.

\section{FIMPs in the early universe}
\label{sec:FIbasics}

To understand the production of FIMPs we first review the underlying
formalism.  The case of FIMP production from decays and scatterings
and their impact on small-scale structures has already been addressed
in several recent
works~\cite{Heeck:2017xbu,Boulebnane:2017fxw,Bae:2017dpt,Ballesteros:2020adh,DEramo:2020gpr}. Nevertheless,
here we briefly summarise the relevant steps of the computation and
precise, where relevant, new inputs compared to previous
literature. We also detail our implementation of FIMP momentum distribution functions in the public Boltzmann code
\class \footnote{Our modified \textsc{class} version can be found at \href{https://github.com/dchooper/class_fisw}{https://github.com/dchooper/class\_fisw}.
}. Complementary discussion on the Boltzmann equations for FI can
be found in \eg~\cite{Belanger:2018ccd,Belanger:2020npe}.

\subsection{Boltzmann equations}
\label{sec:BE}

In order to describe the momentum distribution of FIMPs, one has to
solve the unintegrated Boltzmann equation for the DM phase-space distribution function $f_\chi(t,p)$
\begin{equation}
  \frac{\D f_\chi}{\D t}={\cal C}[f_\chi]
  \label{eq:fcoll}
\end{equation}
where $ \chi$ refers to the DM particle, with $t$ and $p$ the
proper time and momentum, and ${\cal C}$ refers to the collision
terms responsible for FIMP production from the decays or scatterings
of some mother particle $B$. The
number density of any species $i$ can be obtained by integrating out
the distribution function $f_i (t,p)$ as
\begin{equation}\label{eq:numDens}
  n_i= g_i \int \frac{\D^3 p}{(2\pi)^3}f_i (t,p)\,,
\end{equation}
where $g_i$ is the number of degrees of freedom (dof) of the species $i$.  It is usually
appropriate to re-express proper time and momentum in terms of
independent dimensionless variables. In the context of the DM studied
here, the time variable $t$ is  traded with $x={\mref}/{T}$,
where $\mref$ denotes some reference mass (often the mass of the
mother particle $B$ for FIMP production) and $T$ denotes the
temperature of the standard model bath.  The relation between
$x$, or equivalently $T$, and $t$ can be easily obtained when entropy
is conserved, which we will assume throughout this work. In this case,
we have $\D(s a^3)/\D t=0$, where $s$ is the entropy density and $a$ the
scale factor.  As a result, keeping in mind that $s\propto \gs T^3$,
one obtains
\begin{equation}
  \frac{\D\ln T}{\D\ln t}= -{\bar H} \quad {\rm with }  \quad \bar H=\frac{H}{1+1/3\, \D\ln \gs/\D\ln T} \, ,
  \label{eq:barH}
\end{equation}
where $\gs(T)$ denotes the number of relativistic 
dof in the thermal bath of temperature $T$ contributing to the 
entropy, and $H= \D\ln a/\D t$ is the Hubble expansion rate. In a radiation 
dominated era, the Hubble rate reduces to
\begin{eqnarray}
 H&=&\frac{T^2}{M_0(T)} \quad {\rm with} \quad  M_0(T)=\mpl \sqrt{\frac{45}{4 \pi^3 g_*(T)}}\,,
  \label{eq:M0}
\end{eqnarray}
where $\mpl=1.2 \times 10^{19}$ GeV is the Planck mass and $g_*(T)$ denotes the number of relativistic dof in the 
thermal bath of temperature $T$, this time contributing to the radiation 
energy density.

Here we mostly consider scenarios for which $g_*(T),\gs(T)$ are constant 
before FIMP production. As a result,  $\frac{d\ln T}{d\ln t}=-H$ and  it
is convenient to use
\begin{equation}
  x=\frac{m_B}{T} \quad{\rm and}\quad q=\frac{p}{T}
\label{eq:xq-basic}
\end{equation}
as time and momentum-independent variables\footnote{In full
  generality, $q=p/T$, which is not a time-independent variable as the
  temperature scales as $T\propto g_{*S}^{1/3} a^{-1}$ and $p\propto
  1/a$.} and the Boltzmann equation from eq.~(\ref{eq:fcoll}) simply reduces to
\begin{equation}
  x  H \partial_x f_\chi={\cal C}[f_\chi] \, .
  \label{eq:fcollx}
\end{equation}
In App.~\ref{sec:gSx} we discuss the relevant choice of time and
momentum variable for time-varying $g_*,\gs$.

We assume that the initial FIMP abundance is negligible. We use the
compact notation ${\rm in}\to {\rm fin} + \chi $ for the DM particle $\chi$  production
processes, including decays and scatterings. With ``in'' (``fin'') we
refer to an ensemble of initial (final) state particles as a source
for DM production. In this context, the collision term in
eq.~(\ref{eq:fcollx}) reads
\begin{equation}
  {\cal C}[f_\chi]=\frac{1}{2 g_{\chi}E_\chi}\int \Pi_\alpha\frac{ \D^3 p_\alpha}{(2\pi)^32E_\alpha}(2\pi)^4 \delta^4(P_{\rm fin}+ p_\chi-P_{\rm in}) f_{\rm in} (1\pm f_{\rm fin}) (1\pm f_\chi) |{\cal M}|^2_{{\rm in}\to {\rm fin} + \chi}\,.
  \label{eq:coll}
\end{equation}
In this expression the index $\alpha$ runs over all particles in the
initial and final states except for DM, $P_\alpha$ is the sum of the
four-momenta of initial or final state particles for $\alpha={\rm in}$
and ${\rm fin}$, $f_{\rm in}$ refers to the product of the distribution
functions of the initial state particles, and $(1\pm f_{\rm fin})$
is the product of Pauli blocking (with a minus sign) or
Bose-Einstein enhancing (with a plus sign) factors for final state
particles. Furthermore, $|{\cal M}|^2_{{\rm in}\to {\rm fin} + \chi}$ denotes
the amplitude squared {\it summed} over initial and final state
quantum numbers. For concreteness, we will focus here on 2-body decays 
of the form $B\to A \chi$, and $2 \to 2$ scatterings of the form 
$B B'\to A' \chi$ for DM production. As such, we consider a scenario
where $B$ and $\chi$ are odd under a $Z_2$ symmetry that stabilizes DM\@. We will also neglect spin 
statistics effects by taking $(1\pm f_{\rm fin})=1$, see
\eg~\cite{Belanger:2020npe,DEramo:2020gpr,Ballesteros:2020adh} for
some complementary studies.

In this paper we focus on scenarios in which the mother particle is in
kinetic equilibrium while producing the DM and $m_B> m_\chi$. For $B$
in kinetic equilibrium, its distribution function can be written as
(see \eg~\cite{Binder:2017rgn,Gondolo:1990dk} for a discussion)
\begin{equation}
  f_B(x,q)=\frac{Y_B(x)}{Y_B^{\eq}(x)} f_B^{\eq}(x,q)\,,
  \label{eq:fBSW}
\end{equation}
where $f^{\eq}(x,q)$ denotes the usual equilibrium distribution
function with zero chemical potential. In order to derive an analytic
estimate for the DM distribution function, we will consider a Maxwell-Boltzmann distribution for $B$, but we have explicitly checked
numerically that the results do not change significantly when
considering \eg~a Bose-Einstein distribution, see
also~\cite{Heeck:2017xbu,Belanger:2018ccd,DEramo:2020gpr}.

The average DM momentum at the time
of production, and its subsequent redshifted value, provide a good tool to
estimate the importance of cosmological constraints arising from small-scale structure, more specifically the Lyman-$\alpha$ power flux
constraints and the number of extra relativistic dof, see
\eg~\cite{Heeck:2017xbu,Bae:2017dpt,Ballesteros:2020adh,DEramo:2020gpr}
and also \eg~\cite{Baldes:2020nuv} in a slightly different context.   
In particular, the rescaled $n^{\text{th}}$-moment of the distribution is obtained evaluating
\begin{equation}
  \langle q^n\rangle= \frac{\int \D^3 q \, q^n f_\chi(q) }{\int \D^3 q \, f_\chi(q)}\,,
\label{eq:qn}
\end{equation}
where $f_\chi(q)$ is the FIMP distribution after production ($\equiv f_\chi(x,q)$ for $x\gg x_{\rm prod}$).

\subsection{FIMPs from decays}
\label{sec:FI}

For DM production through decays, the collision term in the Boltzmann eq.~(\ref{eq:fcollx}) reduces to
\begin{equation}
  {\cal C}_{\rm dec}[f_\chi]=\frac{x}{16 \pi g_{\chi} q\sqrt{q^2m_B^2 + m^2_\chi x^2}}\int_{\xi_-}^{\xi_+} \D \xi_B f_B |{\cal
    M}|^2_{B\to A\chi} \,,
  \label{eq:Cdec}
\end{equation}
where $\xi_B = E_B/T$ and the values of $\xi_{\pm}$ are discussed in App.~\ref{sec:more}, see also~\citep{Boulebnane:2017fxw}. In what
follows, we distinguish between the FI and the SW production from
decays of a mother particle that is in kinetic equilibrium with the
thermal bath. In the case of FI production, discussed in
Sec.~\ref{sec:FI-dec}, the mother is both in kinetic and chemical
equilibrium.  On the other hand, SW production would refer to the DM
production after $B$ freeze-out, \ie~after $B$ chemically decouples,
see Sec.~\ref{sec:SW}.  Accordingly, the two contributions -- although
stemming from the very same decay process -- can arise at different
times with distinct mean momenta and momentum distributions. This is
illustrated in Sec.~\ref{sec:FI-SW}.  In this context, it is
convenient to introduce the dimensionless ratio
\begin{equation}
R_\Gamma^{\rm prod}=\frac{M_0(T_{\rm
    prod}) \Gamma_{B\to A\chi}}{m_B^2},
  \label{eq:Rgam}
\end{equation}
where $M_0(T_{\rm prod})$ corresponds to the rescaled Planck mass of
eq.~(\ref{eq:M0}) with the number of relativistic dof
estimated at the DM production temperature $T_{\rm prod}$. 

\subsubsection{Freeze-in from decays}
\label{sec:FI-dec}

The largest contribution to DM freeze-in from decays of a bath
particle $B$, arises around $\xFI=m_B/T\sim 3$~\cite{Hall:2009bx} due
to the interplay of two competing effects.  On the one hand, in a
radiation dominated era, $\Gamma_{B\to A\chi}/H$ increases with $x$,
leading the decay to become more efficient at late times.  On the
other hand, once the bath particle becomes non-relativistic,
\ie~$x\gtrsim1$, its number density starts to decrease exponentially.

Considering renormalisable interactions in the radiation dominated era
and assuming\footnote{If the DM mass is not neglected in
  the computation of the DM distribution function, a further analytic expression for the latter would be needed, while an expression for $\partial_x f_\chi$ is given
in~\cite{Boulebnane:2017fxw}. Integrating out the
  distribution function numerically, Ref.~\cite{Boulebnane:2017fxw}
  showed that the analytic form of $f_\chi$ obtained in the limit
  $m_\chi\to 0$, Eq.~(\ref{eq:fdec}), is a very good approximation in the range of $q$ relevant to extract the Lyman-$\alpha$ constraints.}
$m_\chi\ll m_B, m_A$ as well as a Maxwell-Boltzmann distribution for
the mother bath particle $B$, \ie~$f_{B}= \exp(-E_B/T)$, we can obtain
a simple analytic expression for
$f_\chi$ of the form~\cite{Heeck:2017xbu,Bae:2017dpt}
\begin{eqnarray}
 g_{\chi}  f_\chi^\mathrm{FI,\,dec}(q)&=& 2g_B \frac{ R_\Gamma^{\rm FI}}{\delta^3} \sqrt{\frac{\pi \delta}{q}} \exp\left( - \frac{q}{\delta}\right),\label{eq:fdec}\\
  &{\rm with} &\delta=\frac{m_B^2-m_A^2}{m_B^2}\label{eq:deltdec}
\end{eqnarray}
where we use the short-hand notation $f^\mathrm{FI,\,dec}_{\chi}(q)=f^\mathrm{FI,\,dec}_{\chi} (x\to\infty,q)$. Furthermore, $g_B$ is the
number of dof of $B$, $\TFI$ is the temperature at FI production, which
is $\TFI= m_B/x_{\rm FI}$. Further details on the computation and involved approximations are given in App.~\ref{sec:more}.
Integrating out eq.~(\ref{eq:fdec}) over momenta, one obtains the DM abundance from FI,
\begin{equation}
  \Omega_\chi h^2|_\mathrm{FI,\,dec}=m_\chi\times\frac{135}{8\pi^3}\frac{g_B}{g_*\q T_{\rm FI}\w}R^\mathrm{FI}_\Gamma \frac{s_0h^2}{\rho_\mathrm{crit}}\,,
\end{equation}
where $\rho_\mathrm{crit}=3 \mpl^2 H_0^2/(8\pi)$ is the critical energy density,  $s_0$ is the entropy density today, and $h$ is the rescaled Hubble parameter today, $h=H_0/(100\,\text{km}\,\text{s}^{-1}\text{Mpc}^{-1})\sim 0.7$.
Making use of eqs.~(\ref{eq:qn}) and
(\ref{eq:fdec}), the $n^{\rm th}$-moment of the rescaled DM momentum distribution of
(\ref{eq:fdec}) is given by
\begin{equation}
  \qnav|_\mathrm{FI,\,dec}= \frac{4}{3\sqrt{\pi}}\Gamma\left(\frac{5}{2}+n\right)\times \delta^{n}\,,
  \label{eq:qavFIdec}
\end{equation}
where the $\Gamma$ denotes the mathematical Gamma-function.  In
particular, $\qav|_\mathrm{FI,\,dec}=5/2 \times \delta$ while for thermal
WDM one would get $\qav_{\rm thermal}\simeq 3$, see
\eg~\cite{Heeck:2017xbu} for a discussion.

\subsubsection{FIMPs from superWIMP mechanism}
\label{sec:SW}

After the time at which $B$ gets chemically decoupled, usually referred to
as freeze-out time, around $x_{\rm FO}\sim 25$, the frozen out particle
eventually decays into DM and, hence, provides a contribution to the
DM abundance. This DM production mechanism is usually referred to as
the SW mechanism. Interestingly, the associated DM phase-space
distribution might also peak at significantly higher $q$ values than
in the case of FI production.

To get an analytic expression of the DM phase-space distribution, we
employ the ansatz of eq.~(\ref{eq:fBSW}) for the bath particle
distribution, together with the non-relativistic expression for the $B$
equilibrium comoving density, $Y_B^{\eq}(x)$. After chemical decoupling only
late $B$ decays can affect the $B$ abundance so that $Y_B$ should satisfy
\begin{equation}
\frac{\D \ln Y_B}{\D x}=-R^\mathrm{SW}_\Gamma x \frac{K_1(x)}{K_2(x)}\quad \Rightarrow  \quad Y_B(x)\simeq Y_{\rm FO} e^{- R_\Gamma^{\rm SW}(x^2-x^2_{\rm FO})/2 }  \quad[x>x_{\rm FO} ]\,,
  \label{eq:YB}
\end{equation}
where $R^{\rm SW}_\Gamma$ is given by eq.~(\ref{eq:Rgam}) with
$M_0=M_0( \TSW)$ and $Y_{\rm FO}$ is the roughly constant frozen-out
bath particle abundance between $B$ chemical decoupling and complete decay to DM at $x_{\rm
  SW}$, \ie~$Y_B\simeq Y_{\rm FO}$ for $x_{\rm FO}\lesssim x\lesssim x_{\rm
  SW}$.  In order to derive the above analytic expression we have
further assumed that ${K_1(x)}/{K_2(x)} \simeq 1$ in the
non-relativistic limit, as well as a constant number of relativistic
dof\@. From eq.~(\ref{eq:YB}) it is clear that the characteristic temperature parameter at which the decay takes place is
 \begin{equation}
   x_{\rm SW}=\sqrt{\frac{2}{R_\Gamma^{\rm SW}}} \,.
   \label{eq:xSW}
 \end{equation}

 Plugging the above inputs into eq.~(\ref{eq:Cdec}) we can readily
 integrate over the $\xi_B$ with the lower integration bound $\xi_{B\, \rm min}=q/\delta+\delta x^2/(4q)$ and
 get
\begin{eqnarray}
g_{\chi}  \partial_x  f_\chi^{\rm SW}(x,q)&=& \frac{Y_B(x)}{Y_B^{\eq}(x)} \times\frac{g_B }{ \delta}\frac{x^2 }{ q^2} R_\Gamma^{\rm SW} \exp\left( -q/\delta-\delta x^2/(4 q)\right)\,.
 \label{eq:dfXdxR}
  \end{eqnarray}
Integrating eq.~(\ref{eq:dfXdxR}) over $x$ we obtain
\begin{eqnarray}
  g_{\chi}f_\chi^{\rm SW}(q)&\simeq& \sqrt{8\pi} \, \frac{ C_\mathrm{SW} }{q \delta}\, \exp\!\left(-\frac{2 R^{\rm SW}_\Gamma q^2}{\delta^2}\right) \cr
   &{\rm with}& C_{\rm SW}=\gs(x_{\rm SW}) Y_{\rm FO} \frac{R^{\rm SW}_\Gamma}{\delta}(2 \pi)^{3/2}\frac{ 2 \pi^2 }{45 }\,,
  \label{eq:fSW}
\end{eqnarray}
where $\gs$ has to be evaluated at the temperature of SW
decay. 
To derive such a simple
expression, we have assumed that the relevant $(x,q)$ parameter space
for SW corresponds to $x\gg x_{\rm FO}$ and $2 q R^{\rm
  SW}_\Gamma\ll\delta$, see App.~\ref{sec:more} for details. In
addition, the results derived here assumed that $g_{*S}$ is constant throughout
SW production. While this is not always true, we have explicitly checked
that when considering $\gs=\gs(x_{\rm SW})$ in eq.~(\ref{eq:fSW}) the results are in very good agreement with numerical calculations taking a time-dependent $g_{*S}$ into account, see the discussion in App.~\ref{sec:gSx}.\footnote{Notice that in~\cite{Merle:2015oja,Baumholzer:2019twf}, $B$
  has been assumed to be kinetically decoupled since freeze-out time,
  i.e. eq.~(\ref{eq:fBSW}) does not hold. This is usually not the case
  when $B$ is charged under standard model gauge group, which we assume here.  
  Therefore, we cannot directly compare our results to theirs.} 
Finally, integrating out eq.~(\ref{eq:fSW}) over momenta, we simply recover that the DM abundance arising from SW,  $Y_\chi^{\rm SW}$ is equal to $Y_{\rm FO}$, confirming the consistency of our approach.
We can also easily evaluate the 
$n^\mathrm{th}$-moments of the DM rescaled momentum distribution (eq.~(\ref{eq:fSW})) from SW
production, which reduces to 
\begin{equation}
\qnav|_{\rm SW}\simeq  \q 2 R^{\rm SW}_\Gamma\w^{-n/2}\delta^n\,\Gamma\left(\frac{n}{2}+1\right)\, .
\end{equation}
In particular, for $n=1$, we have $\qav|_{\rm SW}=\delta\sqrt{\frac{\pi}{8R^{\rm SW}_\Gamma}}$.

\subsubsection{When superWIMP meets freeze-in  }
\label{sec:FI-SW}

\begin{figure*}[t]
  \begin{center}
    \includegraphics[width=.495\textwidth]{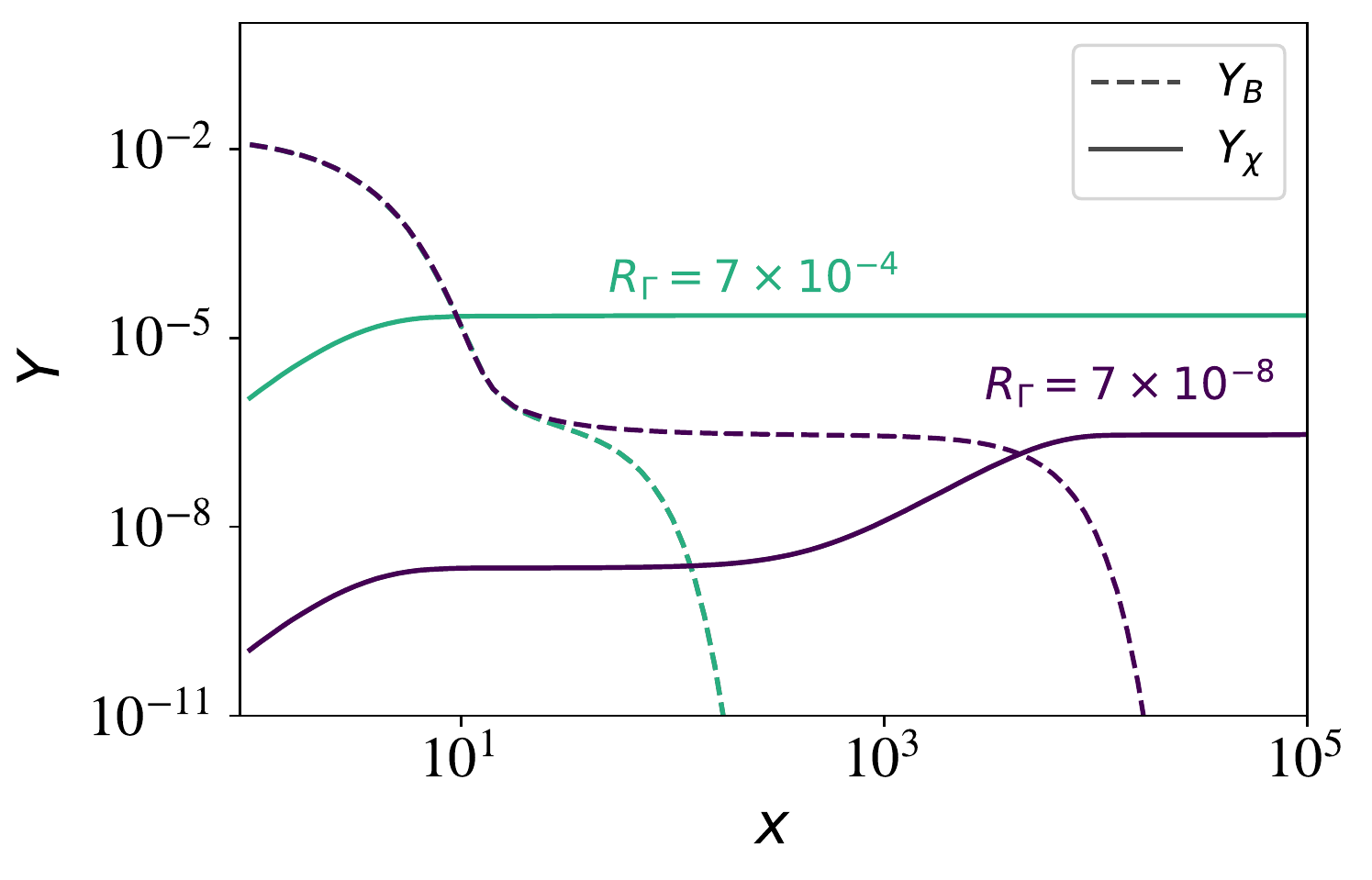} \includegraphics[width=.485\textwidth]{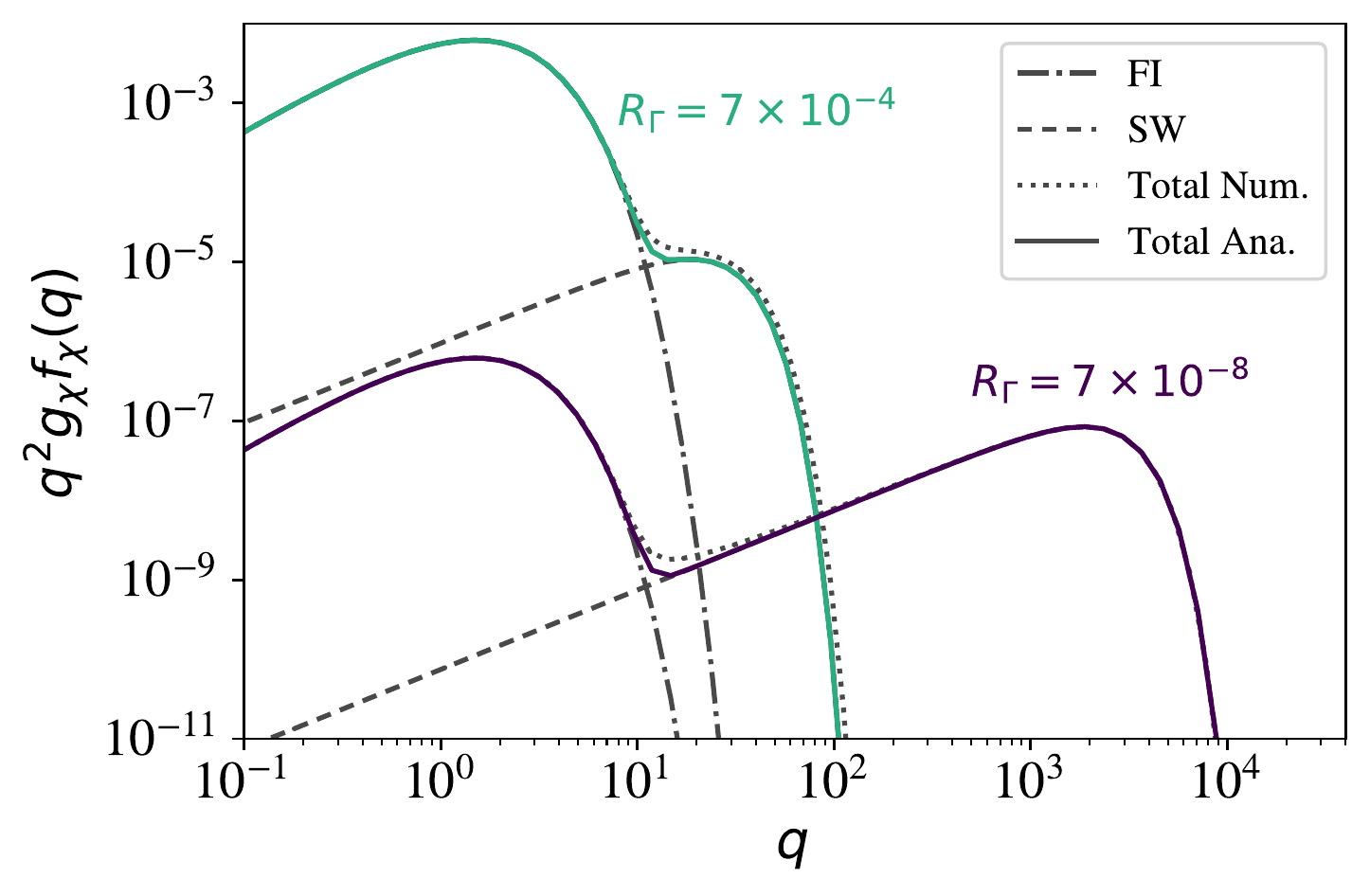}
  \vspace*{-5mm}
  \end{center}
  \caption{FIMP production from $B$ decays with $B$ in kinetic
    equilibrium with the standard model bath.  Two benchmarks are displayed taking
    $R_\Gamma= 7\times 10^{-4}$ (green curves) and $R_\Gamma= 7\times
    10^{-8}$ (purple curves). \textbf{Left}: Bath particle
    (dashed curves) and DM (solid curves) comoving number density 
    as a function of the time variable $x$.  \textbf{Right}: FIMP
    distribution function multiplied by the momentum squared, $q^2
    g_{\chi}f_\chi(q)$, as a function of the rescaled momentum $q$. The
    analytic FI and SW contributions are shown with grey dashed and
    dot dashed curves respectively while the coloured solid lines
    correspond to the sum of the latter two. With the grey dotted
    curves we also show the results obtained by
    integrating eq.~(\ref{eq:Cdec}) without any approximation.  }
\label{fig:PSD}
\end{figure*}

As mentioned above, one single decay process can give rise to two types
of FIMP DM production mechanisms: one from FI and another
from SW. In Fig.~\ref{fig:PSD} we illustrate the comoving number
densities evolution as a function of the temperature parameter $x$ (left), and the DM distribution
function $f_\chi (q)$ dependency in rescaled momentum (right) for two
benchmarks taking $R_\Gamma= 7\times 10^{-4}$ (green curves) and
$R_\Gamma= 7\times 10^{-8}$ (purple curves). We have assumed
$m_B\gg m_A$, such that $\delta =1$ and $g_*=g_{*,S}= 106.75$ at both
$T_{\rm FI}$ and $T_{\rm SW}$, \ie~$R_\Gamma=R_\Gamma^{\rm FI}=R_\Gamma^{\rm SW}$.

In the left panel of Fig.~\ref{fig:PSD}, we show both $Y_B(x)$, the
bath particle comoving abundance (dashed lines), and
$Y_{\chi}(x)$, the DM comoving abundance (solid lines). At early times,
$Y_B$ follows the equilibrium Maxwell-Boltzmann distribution which is already
becoming exponentially suppressed around $x\sim 1$. At chemical
decoupling, for $x=x_{\rm FO}$, $Y_B$ freezes-out and remains
constant, with $Y_B=Y_{\rm FO}$, up until $x\sim x_{\rm SW}$ where it fully
decays to DM\@. In parallel, the DM abundance is slowly
produced up until $x_{\rm FI}\sim 3$ where it freezes in at a value
$Y_{\chi} (x_{\rm FI})$.  The second contribution to the DM abundance
from the SW mechanism is produced around $x_{\rm SW}\sim 53$
  and $5.3\times 10^4$ for $R_\Gamma= 7\times 10^{-4}$ and $R_\Gamma= 7\times
10^{-8}$, respectively, contributing around 2 \% and 99\% to the
relic DM abundance.  If $Y_B(x_{\rm FO})$ is large enough
compared to $Y_\chi(x_\mathrm{FI})$, the SW contribution can significantly
affect the DM abundance, as visible for $R_\Gamma= 7\times 10^{-8}$
(purple curve). 

In the right panel of Fig.~\ref{fig:PSD}, we show the DM distribution
multiplied by the rescaled momentum squared, $q^2 g_\chi f_\chi (q)$, as a function of
$q$. The FI from $B$ decay contribution to $f_\chi (q)$, as in
eq.~(\ref{eq:fdec}), is shown with grey dot dashed curves while the SW
contribution from eq.~(\ref{eq:fSW}) is shown with dashed curves. The
sum of the latter two analytic results is shown with coloured solid
curves. For comparison, we show with grey dotted curves the
numerical result obtained integrating out the collision term of
eq.~(\ref{eq:Cdec}) without any approximations.  We see that both
coloured and grey dotted lines give rise to very similar results. More
quantitatively, for $R_\Gamma= 7\times 10^{-4}$ ( $7\times 10^{-8}$)
we introduce a relative error below 1\% (around 2\%)
in estimating the DM relic
abundance by integrating out the analytic result instead of the
numeric result. This, in particular, illustrates that the analytic
results derived in the previous section provide a very good estimate
of the SW contribution to DM abundance and distribution function. From
this figure, it is also clear that the FI through decay distribution
peaks around $q\sim {\cal O}(1)$ as expected from
eq.~(\ref{eq:qavFIdec}) while the SW distribution is always expected
to peak at larger $q$ values giving rise to a multimodal DM
distribution.

As a final comment, let us also mention that while here we
illustrate the case where FI and SW contributions arise from the same
mother particle, $B$, the most relevant contribution to each
production mechanism could also originate from two different
particles, see \eg~\cite{Baumholzer:2019twf}.

\subsection{FIMPs from scatterings}
\label{sec:FI-scat}

FIMPs could also have been produced in the early universe through FI
from scatterings.  In the case of $B B'\to A' \chi$ scatterings,
assuming a Maxwell-Boltzmann distributions for the bath particles $B$ and
$B'$, we have\footnote{In eq.~(\ref{eq:Cscat}), we have an extra
  factor of $1/2$ compared to~\cite{Heeck:2017xbu}, which we believe
  to be a typo, as our numerical integration fully agrees with the
  results of \cite{Bae:2017dpt}.}
\begin{equation}
  {\cal C}_{\rm scat}[f_\chi]=\frac{1}{32\pi^2 g_{\chi} E p}\int_{s_\mathrm{min}} \D s \int_{E_{A'}^\mathrm{min}}\D E_{A'} \, \exp\left(-\frac{E+E_{A'}}{T}\right)\frac{\hat\sigma}{2}\frac{s}{ \sqrt{(p\cdot p_{A'})^2- (m_\chi m_{A'})^2}} \,,
  \label{eq:Cscat}
\end{equation}
where $\hat\sigma(s)$ denotes the reduced $B B'\to A' \chi$
cross-section, which is a function of the centre of mass energy squared $s$, satisfying
\begin{equation}
  \frac{\D \hat\sigma}{\D t}= \frac{1}{8 \pi s} |{\cal M}|^2 \,,
  \label{eq:sighat}
\end{equation}
where the derivative is taken with respect to the Mandelstam variable
$t$, and $|{\cal M}|^2$ is again the transition amplitude squared
summed over initial and final state dof. Going to the limit of
$m_\chi\ll m_A',m_B,m_B'$, eq.~\eqref{eq:Cscat} reduces to 
\begin{eqnarray}
  g_{\chi}f_\chi^\mathrm{FI,\,scat}(q)&=& \frac{1}{32 \pi^2 q^2}\frac{M_0^\mathrm{FI}}{\mref} \int_0^\infty \D x\int_{\tilde s_\mathrm{min}}^\infty \D \tilde s\frac{\hat \sigma \tilde s}{\tilde \Delta}\exp\left( - \frac{q\tilde s}{\tilde\Delta}-\frac{\tilde\Delta}{4 q}\right)\label{eq:fscat}\\
  &{\rm with} &\Delta=s-m_{A'}^2 \quad {\rm and} \quad \tilde\Delta=\Delta/T^2 \,, \label{eq:deltscat} 
\end{eqnarray}
  where we again use the short-hand notation $f_\chi^{\rm
    scat}(q)=f_\chi^\mathrm{FI,\,scat} (x\to\infty,q)$, which is the FIMP
  distribution today when produced through $2\to 2$ scatterings, in
  agreement with~\cite{Heeck:2017xbu}.  In eq.~(\ref{eq:fscat}), we
  denote with a tilde dimensionless variables rescaled with
  temperature with \eg~$\tilde s = s/T^2$.

  As the details of the distribution function from FI through
  scatterings is quite model-dependent, see \eg~\cite{Bae:2017dpt},
  we leave for Sec.~\ref{sec:topphilic} a more thorough discussion on
  the latter in the context of a top-philic DM scenario.  Nevertheless,
  when $s_\mathrm{min} $ and $\hat \sigma$ can be assumed to be temperature-independent, it is possible to get a generic expression for
  $\qnav$ from eq.~(\ref{eq:qn}), namely
 \begin{equation}
 \qnav|_\mathrm{FI,\,scat}= \frac{4}{3\sqrt{\pi}}\Gamma\left(\frac{5}{2}+n\right)\times \left[1+\frac{\int \D s \, \hat \sigma \, \q -1+\q 1-m_{A'}^2/s\w^n\w/s^{3/2}}{\int \D s \, \hat \sigma/s^{3/2}}\right]\,,
\label{eq:qnavsc}
\end{equation}
where the integrals over $s$ run from $s_\mathrm{min}= {\rm max}\q\q
m_B+m_{B'}\w^2, m^2_{A'}\w$ to $\infty$.\footnote{In general,
  in eq.~(\ref{eq:fscat}), the lower integration limit on the centre
  of mass energy squared and the reduced cross-section could be
  explicit functions of the bath temperature, \ie~$s_\mathrm{min} = s_\mathrm{min}\q
  T\w \,$ and $\hat \sigma = \hat \sigma(s,T)$. This is, for example,
  the case when taking into account thermal corrections such as a temperature-dependent mass. In that case the results and implications of eqs.~\eqref{eq:qnavsc} and 
  \eqref{eq:relicDensScat} do not apply.}
The overall prefactor is nothing but $\qnav|_\mathrm{FI,\,dec}$ in the
$\delta=1$ case. In addition, the second term in the squared
parenthesis vanishes when $m_{A'}$ is small with respect to one of the
masses of the initial bath particles. Therefore, it is apparent that,
when there is one initial state particle that is much heavier than the
final state particles, the squared parenthesis in
eq.~(\ref{eq:qnavsc}) reduces to 1, and we recover the FI through decay
result. This in particular implies that FI through decay and scattering distributions share the same $q$-dependence,
\begin{equation}\label{eq: generalPSDScats}
  f_\chi(q)|_\mathrm{FI,\,scat}\propto q^{-1/2} \exp(-q)\,, \qquad [m_{A'},m_\chi\ll m_{B} \text{ or }m_{B'}] \,,
\end{equation}
which would agree with the distributions used in~\cite{DEramo:2020gpr}.
Even when $m_{A'}$ is non-negligible, since $m_{A'}^2\leq s_\mathrm{min}$, the second term in the
squared parenthesis is always negative. We thus find that
\begin{equation}
  \qnav|_\mathrm{FI,\,scat}\leq \frac{4}{3\sqrt{\pi}}\Gamma\left(\frac{5}{2}+n\right)
\label{eq:qnavsclim}
\end{equation}
both for FI from scatterings and from decays. Finally, the contribution to the relic density from FI through  scattering is given by
\begin{align}\label{eq:relicDensScat}
\Omega_\chi h^2|_\mathrm{FI,\,scat} =
 m_\chi\times \frac{s_0}{\rho_\mathrm{crit}/h^2}\frac{135M_0}{256\pi^5g_*(T_\mathrm{FI})}
\times\q  \int_{s_\mathrm{min}}^{\infty}\D s \ph  \frac{\hat \sigma}{s^{3/2}}\w,
\end{align}
assuming again that $s_\mathrm{min} $ and $\hat \sigma$ are
temperature-independent.

\subsection{FIMP distribution functions in {\mdseries \textsc{class}}}
\label{sec:distrib}

In order to precisely follow the cosmological evolution of the FIMPs,
we have implemented the FIMP distribution functions in the public
Boltzmann code \class~\cite{Lesgourgues:2011rh}. For that purpose, it
is convenient to introduce a new rescaled momentum variable,
\begin{eqnarray}
 q_\star= \frac{p (t)}{T_\star(t)} \quad { \rm with}\quad T_\star (t)= c_\star T_\gamma(t_{\rm prod})  
 \frac{\aprod}{a(t)}
 \label{eq:TNCDM_gen}
\end{eqnarray}
where $p\propto 1/a$ is the proper momentum, $c_\star$ is a constant
factor that will be chosen for each FIMP production mode, $\aprod$ and
$T_\gamma(t_{\rm prod})$ are the scale factor and the photon
temperature at the time of production.  The definition of $T_\star
(t)$ is introduced in \class~through the input variable ${\tt
  T_{ncdm}}$ which corresponds to the  ratio of
temperatures $T_\star$ and $T_\gamma$ today.
Using eq.~(\ref{eq:TNCDM_gen}), the latter dimensionless variable
takes the form
\begin{equation}
{\tt T_{ncdm}}=\frac{T_\star(t_0)}{T_\gamma(t_0)}=c_\star  \aprod \frac{ T_\gamma(t_{\rm prod})  
 }{ T_\gamma(t_0)}=c_\star\left(\frac{\gtod}{ \gprod}\right)^{1/3} \,,
\label{eq:TNCDMqCLASS}
\end{equation}
where $t_0$ refers to the time today, the scale factor today is
$a_0=1$ and $\gtod=3.91$. We see that ${\tt T_{ncdm}}$ reduces to the
ratio of relativistic dof at production time and today to the power
1/3 up to the constant prefactor $c_\star$, see
\eg~\cite{Lesgourgues:2011rh,Ballesteros:2020adh} for other NCDM
models.

 In practice, for our implementation of FI and SW in \class, we have
 chosen the $c_\star$ prefactors in eq.~(\ref{eq:TNCDMqCLASS}) to be
 $c^{\rm FI}_\star=\delta$ and $c^{\rm SW}_\star= \delta/\sqrt{2
   R^{\rm SW}_\Gamma}$. This implies that the
 distribution functions for FI from decay and SW of
 eqs.~(\ref{eq:fdec}) and~(\ref{eq:fSW}) take the following simpler
 forms:
\begin{equation}
  \begin{dcases*}
g_{\chi}  f_\chi^\mathrm{FI,\,dec}(q_\star)= 2g_B \frac{ R_\Gamma^{\rm FI}}{\delta^3} \sqrt{\pi}   \times \left[\frac{1}{q_\star^{1/2}}\, \exp\!\left(-q_\star\right)\right]\, & for FI through decays,\\
g_{\chi} f_\chi^{\rm SW}(q_\star)= \frac{4 \sqrt{\pi R_\Gamma^{\rm SW}} C_{\rm SW} }{ \delta^2} \times \left[\frac{1}{q_\star}\, \exp\!\left(-q_\star^2\right)\right]\, & for SW\,,
    \end{dcases*}
\label{eq:fCL}
\end{equation}
where the superscript FI or SW in $R_\Gamma$ reminds that the number of
relativistic dof in $M_0$ have to be determined at $\TFI$ or
$\TSW$. The resulting dimensionless variables ${\tt T_{ncdm}}$
which are provided as an input to the \class~code then read
\begin{equation}
  \begin{dcases*}
     {\tt T_{ncdm}^\mathrm{FI}}= \delta \times \left(\frac{\gtod}{\gs(\TFI)}\right)^{1/3} & and $\TFI= \frac{1}{3} m_B$, \\
     {\tt T_{ncdm}^\mathrm{SW}}= \frac{\delta}{ \sqrt{2 R_\Gamma^{\rm SW}}}\times \left(\frac{\gtod}{\gs(\TSW)}\right)^{1/3} & and   $\TSW= \sqrt{\frac{R_\Gamma}{2} }m_B$.
   \end{dcases*}
   \label{eq:TNCDM}
\end{equation}
Notice that the momentum dependence of the SW distribution in
eq.~(\ref{eq:fCL}) is the same as in the case of moduli decay in a
radiation dominated era, considered in~\cite{Ballesteros:2020adh}. Let us also
mention that in the case of FI through scatterings and under the same
assumptions used to derive eq.~(\ref{eq:qnavsc}), we expect
a similar $q_\star$ dependence as in the case of FI
through decays, but the prefactor would become cross-section dependent
instead of decay-rate dependent, see Sec.~\ref{sec:FI-scat} for
details. Finally, using the above parametrisation in
eq.~(\ref{eq:qn}), the mean rescaled momenta $\langle q_\star\rangle$ and 
the mean rescaled squared momenta $\langle q_\star^2\rangle$ reduce to
\begin{equation}
   \begin{dcases*}
\langle q_\star\rangle_{\rm FI,\,dec}=\frac{5}{2}, \quad \langle q_\star^2\rangle_{\rm FI,\,dec} = \frac{35}{4} & for FI through decays,\\
\langle q_\star\rangle_{\rm SW}=\frac{\sqrt{\pi}}{2},  \quad \langle q_\star^2\rangle=1 & for SW.\\
   \end{dcases*}
   \label{eq:qstarav}
\end{equation}
Following the discussion in Sec.~\ref{sec:FI-scat}, we can just replace the equality sign with $\lesssim$ in the case of FI through scatterings.

We will now use the different quantities introduced in this subsection
in order to characterise the typical NCDM cosmological imprint of FIMP DM
and the associated constraints in the next section.

\section{Imprint of FIMPs on cosmological observables}
\label{sec:cosmo}

Once FIMPs have been produced at a time where the standard model bath temperature
is $T=\Tprod$, with $\Tprod= \TFI$ ($\TSW$) for production from the
FI (SW) mechanism, the resulting DM particles free-stream. If their
velocity is sufficiently large at late times, they can free-stream
from overdense to underdense regions and prevent small-scale structure
formation. Furthermore, if FIMPs are still relativistic at Big Bang
Nucleosynthesis (BBN) or Cosmic Microwave Background (CMB) times, they
constitute extra radiation dof that might be constrained by $\Delta
\Neff$ bounds.

In Secs.~\ref{sec:lyman-alpha}
and~\ref{sec:delta-neff} we study the resulting constraints on cosmological observables. We show that when the DM abundance 
  $\Omega_\chi h^2=0.12$ results at 100 \% from the FI or from the SW mechanism, Lyman-$\alpha$ data provide a lower bound on the DM mass of the form
  \begin{equation}
m_\chi  \gtrsim  \begin{dcases*}
 m_{\rm FI}^{\rm lim} \times \delta  \times \left( \frac{106.75}{\gs(\TFI)} \right)^{1/3}& for FI through decays,\\
 m_{\rm SW}^{\rm lim}\times \delta \times \left( \frac{106.75}{\gs(\TSW)} \right)^{1/3} \times \left(R_\Gamma^{\rm SW}\right)^{-1/2} & for SW,\\
   \end{dcases*}
   \label{eq:limsly}
\end{equation}
where the prefactors $m_{\rm FI,SW}^{\rm lim}$ are in the keV mass
range, see the summary in Tab.~\ref{tab:NCDMnounds}.  The results for
FI are valid for FI from decays as well as for any FI from scattering
scenario that would give rise to an equality in
eq.~(\ref{eq:qnavsclim}).  Our results for FI are in very good
agreement with the previous literature
in~\cite{Boulebnane:2017fxw,Bae:2017dpt,Ballesteros:2020adh,DEramo:2020gpr}
when using the same methodology,\footnote{Let us in particular
  emphasise that for the fit to the power spectrum and the area
  criterion, our results are obtained by switching the perfect fluid
  approximation off in \class, which is the only valid approximation
  for generic NCDM, see the discussion in App.~\ref{sec:fit_fluid}.}
see also~\eg~\cite{Dvorkin:2019zdi,Dvorkin:2020xga} for similar
results obtained in a slightly different context.  On the other hand,
for mixed FI-SW scenarios a more detailed analysis is needed, see
Sec.~\ref{sec:mixed}.

\renewcommand{\arraystretch}{1.15}
\begin{table}[t]
	\centering
	\begin{tabular}{ | c | c | c | c | }
		\hline
		Probe & NCDM test & $m_{\rm FI}^{\rm lim}$ [keV] & $m_{\rm SW}^{\rm lim}$ [keV]  \\ 
		\hline \hline
		\multirow{3}{2cm}{\centering Lyman-$\alpha$}&Velocity dispersion, Sec.~\ref{sec:approx-lyman-al} & 16 & 3.8 \\
                  &Fits to transfer function, see Sec.~\ref{sec:pure-fi-sw} & 15 & 3.9 \\
	       &Area criterion, see Sec.~\ref{sec:mixed} & 15 & 3.8\\
		\hline
		$\Delta \Neff$ & see Sec.~\ref{sec:delta-neff}
                & $1.3\times 10^{-2}$& $3.4\times 10^{-3}$\\
			\hline
	\end{tabular}
	\caption{ Mass scales in keV entering into the lower bounds of
          the FIMP masses of eq.~(\ref{eq:limsly}). They arise from
          the FIMP NCDM imprint on cosmological structures assuming that
          100\% of the DM content results from FI or SW mechanism
          production. The values for $m_{\rm FI,SW}^{\rm lim}$
          correspond to the WDM bounds $m_{\rm WDM}^{\mathrm{Ly}\alpha}>5.3$
          keV and $\Delta\Neff (T_{\rm BBN})<0.31$.}
	\label{tab:NCDMnounds}
\end{table}

\subsection{FIMP free-streaming and Lyman-$\alpha$ bound}
\label{sec:lyman-alpha}

The Lyman-$\alpha$ forest flux power spectrum probes hydrogen
clouds at redshifts $2\lesssim z \lesssim 6$. It
provides constraints on the matter power spectrum on small
scales~\cite{Ikeuchi1986, Rees1986}. The scales tested by
Lyman-$\alpha$ data, typically 0.5 Mpc/h $ < \lambda <$ 100
Mpc/h~\cite{Murgia:2017lwo}, are in the  non-linear regime so
that computationally expensive hydrodynamical N-body simulations would
be required in order to properly test a given NCDM scenario. These
expensive simulations have been performed for thermal WDM\@. Following
the early work of~\cite{Viel:2013fqw}, the analysis of
\cite{Palanque-Delabrouille:2019iyz} obtained a bound of $m_{\rm
  WDM}^{\mathrm{Ly}\alpha}=5.3\,$keV at $95\,\%$ confidence level (CL) from Lyman-$\alpha$ flux
observations. It has, however, been argued that the assumptions made
about the instantaneous temperature and pressure effects of the
intergalactic medium in this work might have been too strong.
Relaxing these assumptions \cite{Garzilli:2019qki} found a bound of
$m_{\rm WDM}^{\mathrm{Ly}\alpha}=1.9\,$keV at $95\,\%$ CL\@. We take the latter
as a conservative bound on the thermal WDM mass while the one
of~\cite{Palanque-Delabrouille:2019iyz} will be considered as a
stringent bound.

To circumvent the need for new N-body simulations for these models, in
this paper we implement the FIMP distribution functions discussed in 
Sec.~\ref{sec:distrib} in the Boltzmann code \class. We use this to
extract the linear matter power spectrum of our NCDM scenarios, as well as  
the corresponding transfer functions discussed in Sec.~\ref{sec:pure-fi-sw}.  We then follow a strategy similar to those applied to NCDM in
\eg~\cite{Murgia:2017lwo,Bae:2017dpt,Baldes:2020nuv,Ballesteros:2020adh,DEramo:2020gpr}. In Secs.~\ref{sec:approx-lyman-al}
and~\ref{sec:pure-fi-sw} we extract a lower bound on the DM mass in pure FI
and SW scenarios, making use of the DM velocity dispersion and of
fits to the transfer functions.  Notice that these
constraints are only valid for FIMPs accounting for 100\% of the DM
content. In Sec.~\ref{sec:mixed}, we address the case of the mixed
FI-SW scenarios, or equivalently cases where a given production
mechanism cannot account for all the DM, by applying the area
criterion introduced in~\cite{Murgia:2017lwo}.

\subsubsection{Velocity dispersion }
\label{sec:approx-lyman-al}

If the DM distribution is simple, \eg~with one local maximum, one can
expect that an estimate of the bound on the FIMP mass can be derived
by comparing the typical velocity of the NCDM candidate to the one of
the thermal WDM for which dedicated hydrodynamical simulations have
been performed.  Here we follow the same approach as the one
  proposed by~\cite{Bae:2017dpt}, where an estimated Lyman-$\alpha$
  bound was obtained by considering the root mean square (rms)
  velocity of DM today, $ \sqrt{\langle p^2\rangle_0}/m_\chi$. Here
  $\langle p^2\rangle_0$ refers to today's second moment of the
  momentum distribution, directly related to the velocity dispersion
  of the DM today. When DM arises from one single
  production mechanism or production channel $\sqrt{\langle
    p^2\rangle_0}/m_\chi= \sqrt{\langle q_\star^2\rangle}{\tt
    T_{ncdm}}/m_\chi T_\gamma(t_0)\,$. The lower bound
\begin{equation}
  m_\chi\gtrsim 1.75 \,{\rm keV} \times \sqrt{\langle q_\star^2\rangle}{\tt T_{ncdm}}\times\left(\frac{m_{\rm WDM}^{\mathrm{Ly}\alpha}}{\rm keV}\right)^{4/3}
  \label{eq:Kamada}
\end{equation}
is obtained imposing that the rms velocity, $ \sqrt{\langle
  p^2\rangle_0}/m_\chi$, computed for a FIMP of mass $m_\chi$ equals
the rms velocity for a thermal WDM candidate of mass $m_{\rm
  WDM}^{\mathrm{Ly}\alpha}$ saturating the \mbox{Lyman-$\alpha$}
bound. Notice that $\sqrt{\langle q_\star^2\rangle}$ in
eq.~(\ref{eq:Kamada}) corresponds to the warmness parameter $\tilde
\sigma$ of~\cite{Bae:2017dpt} and that~\cite{Ballesteros:2020adh}
derived the same constraints by equating the equation of states of the
FIMP and the WDM following the early work of~\cite{Colombi:1995ze}.
Eq.~(\ref{eq:Kamada}) was also used in~\cite{DEramo:2020gpr} in the
context of FI to be compared to other methodologies. In those
references it has already been argued that eq.~\eqref{eq:Kamada} can
provide a very good estimate of the Lyman-$\alpha$ constraint for
FIMPs\@. Additionally, in~\cite{Jedamzik:2005sx} the DM velocity is computed in order to derive constraints on the WDM arising from the SW mechanism, and perfectly agrees with the rms
  velocity used here to extract Lyman-$\alpha$ constraints. 
  Using the stringent WDM limit $m_{\rm
  WDM}^{\mathrm{Ly}\alpha}=5.3$ keV
from~\cite{Palanque-Delabrouille:2019iyz}, the Lyman-$\alpha$ bound on
FIMP DM of eq.~(\ref{eq:Kamada}) gives the lower bound on the DM mass
reported in eq.~(\ref{eq:limsly}) with $m_{\rm FI}^{\rm lim}= 16$ keV
and $m_{\rm SW}^{\rm lim}= 3.8$ keV, as given in the first line of
Tab.~\ref{tab:NCDMnounds}.  When using the conservative bound of
$m_{\rm WDM}^{\mathrm{Ly}\alpha}=1.9$ keV from \cite{Garzilli:2019qki}
the prefactors in eq.~(\ref{eq:limsly}) reduce to $m_{\rm FI}^{\rm
  lim}=4.0$ keV and $m_{\rm SW}^{\rm lim}= 0.97$ keV.

In the cases where NCDM would only account for part of the DM content
a dedicated analysis should be performed to compare to the case of
thermal WDM~\cite{Baur:2017stq}. However, as suggested
in~\cite{Bae:2017dpt}, when multiple  production
channels are at the origin of the DM relic abundance but the
total DM distribution is unimodal, one can still use the
rms velocity  ${\sqrt{\langle p^2\rangle_0}}/{m_\chi}$  to extract a bound on the
DM mass. Considering the definition of the second moment of
the momentum distribution, it can be shown that
\begin{equation}
\frac{\sqrt{\langle p^2\rangle_0}}{m_\chi}=\frac{ T_\gamma(t_0)}{m_\chi}\left(\sum_{\rm prod} \left(\frac{\Omega_\chi h^2|_{\rm prod}}{\Omega_\chi h^2}\right) \times \left(\langle
  q_\star^2\rangle {\tt T_{ncdm}^2}\right)|_{\rm prod}\right)^{1/2}\,,
\label{eq:rms-mix}
\end{equation}
where the sum runs over the FIMP production mechanisms,
$\Omega_\chi h^2|_{\rm prod}$ refers to the $\chi$ relic abundance from a
given production channel while $\Omega_\chi h^2$ refers to the total relic abundance. A
first naive estimate of the Lyman-$\alpha$ bound in the case of mixed
scenarios could thus be extracted by comparing the quantity $\sqrt{\langle p^2\rangle_0}/m_\chi$ to the
one of thermal WDM saturating the Lyman-$\alpha$ bound when
$\Omega_\chi h^2=0.12$. Within this framework, we get
\begin{equation}
  m_\chi\gtrsim 1.75 \,{\rm keV} \times \left(\frac{m_{\rm WDM}^{\mathrm{Ly}\alpha}}{\rm keV}\right)^{4/3} \times \left[\sum_{\rm prod} \left(\frac{\Omega_\chi h^2|_{\rm prod}}{0.12}\right) \times \left(\langle
  q_\star^2\rangle {\tt T_{ncdm}^2}\right)|_{\rm prod}\right]^{1/2} \,,
  \label{eq:Kamada-mix}
\end{equation}
where it has been assumed that $\Omega_\chi h^2=0.12$ in order to
compare to the thermal WDM constraints.  Let us emphasise that
eq.~(\ref{eq:Kamada-mix}) is only valid if the total FIMP
distribution, arising from different production processes, is
unimodal. This is, for example, the case of FIMPs from FI through
scatterings and decays analysed in \eg~\cite{Bae:2017dpt}. When the DM distribution is multimodal, as \eg~in a mixed FI-SW scenario,
the area criterion introduced in~\cite{Murgia:2017lwo} should be used instead, see
the discussion in Sec.~\ref{sec:mixed} below.

\subsubsection{Fits to transfer function}
\label{sec:pure-fi-sw}

In order to parametrise the small-scale suppression of the matter
power spectrum within a given NCDM model with respect to the
equivalent CDM case, one can express the ratio between the CDM power
spectrum, $P_{\rm{CDM}}(k)$, and the power spectrum of some new DM
species $X$, $P_{X}(k)$, in terms of the transfer function $T_X$,
defined as
\begin{equation}
P_{X}(k) = P_{\rm{CDM}}(k) \, T^2_{X}(k) \,,
\label{eq:pwdm}
\end{equation}
where $k$ is the wavenumber. It has been shown that the transfer function for 
some NCDM scenarios can be parametrised in terms of a finite set of parameters and
physical inputs.

In particular, in the thermal WDM case,~\cite{Bode:2000gq,Viel:2005qj} use the following parametrisation to
describe the transfer function,
\begin{equation}
T_{X}(k) = \left(1+ (\alpha_{X} k)^{2\mu}\right)^{-5/\mu} \,,
\label{eq:twdm}
\end{equation}
where $\mu$ is a dimensionless exponent and $\alpha_X$ is the breaking
scale. A more general parametrisation that can be applied to a larger
set of NCDM models was also introduced
in~\cite{Murgia:2017lwo,Murgia:2018now,Archidiacono:2019wdp}.

In the case of thermal WDM,~\cite{Viel:2005qj} obtained a very 
good fit for $\alpha$ and $\mu$ from dedicated N-body simulations. We 
will make use of this fit, but with a minor modification to the 
numerical prefactor motivated in App.~\ref{sec:fit_fluid}, where we 
also discuss the validity of this prescription. As such, the breaking scale we will use for
eq.~(\ref{eq:twdm}) is given by $\mu=1.12$ and
\begin{eqnarray}
  	\alpha_{\rm{WDM}} 
		&=&  0.045\q \frac{m_{\rm WDM}}{1\,\text{keV}}\w^{-1.11}\q\frac{\Omega_{\rm WDM}}{0.25}\w^{0.11}\q\frac{h}{0.7}\w^{1.22}h^{-1}\text{Mpc}\,,\label{eq: alphaWDM}	
\end{eqnarray}
in terms of the WDM mass $m_{\rm WDM}$. 

\begin{figure*}[t]
  \begin{center}
    \includegraphics[width=.48\textwidth]{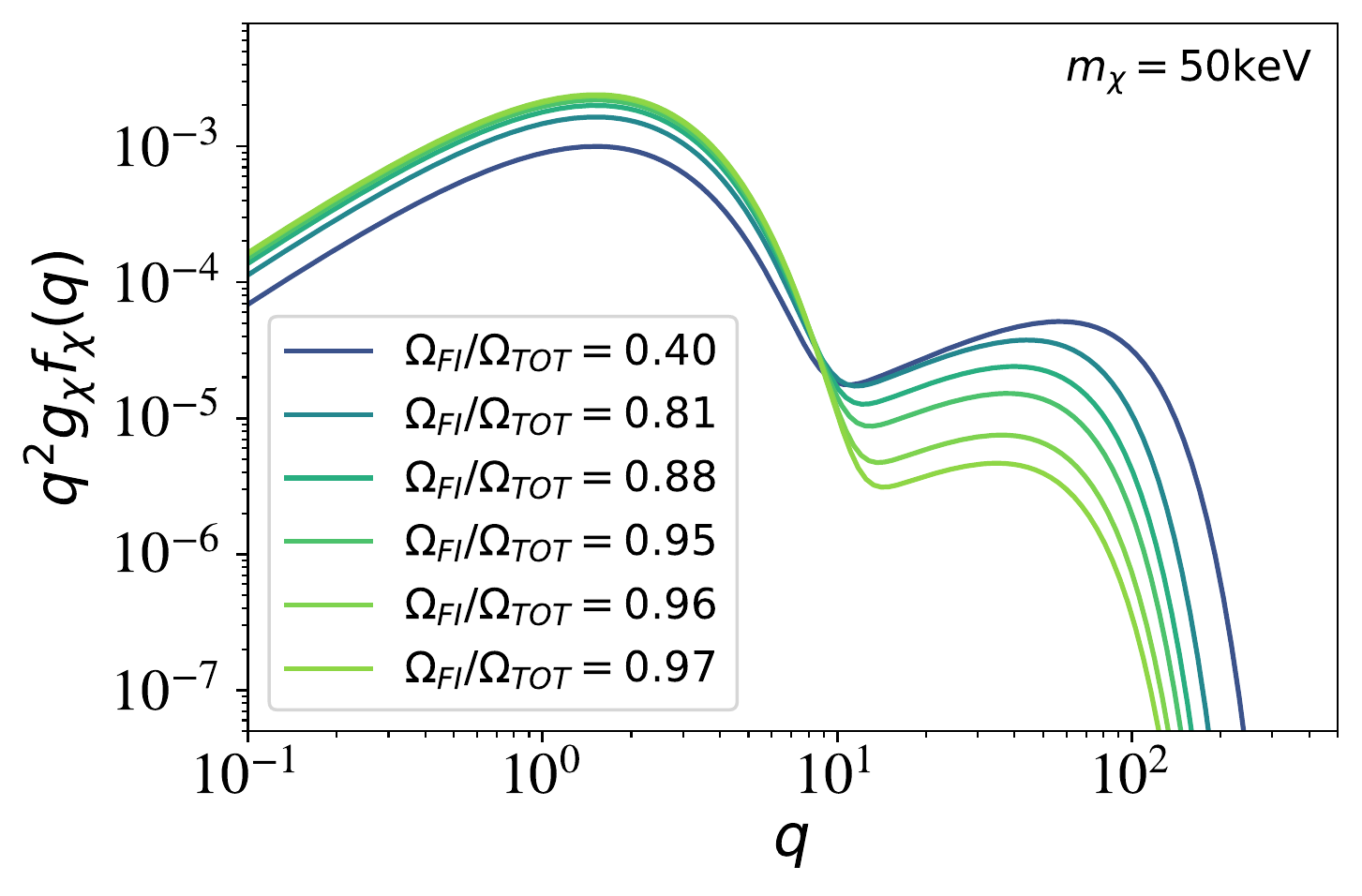} \includegraphics[width=.45\textwidth]{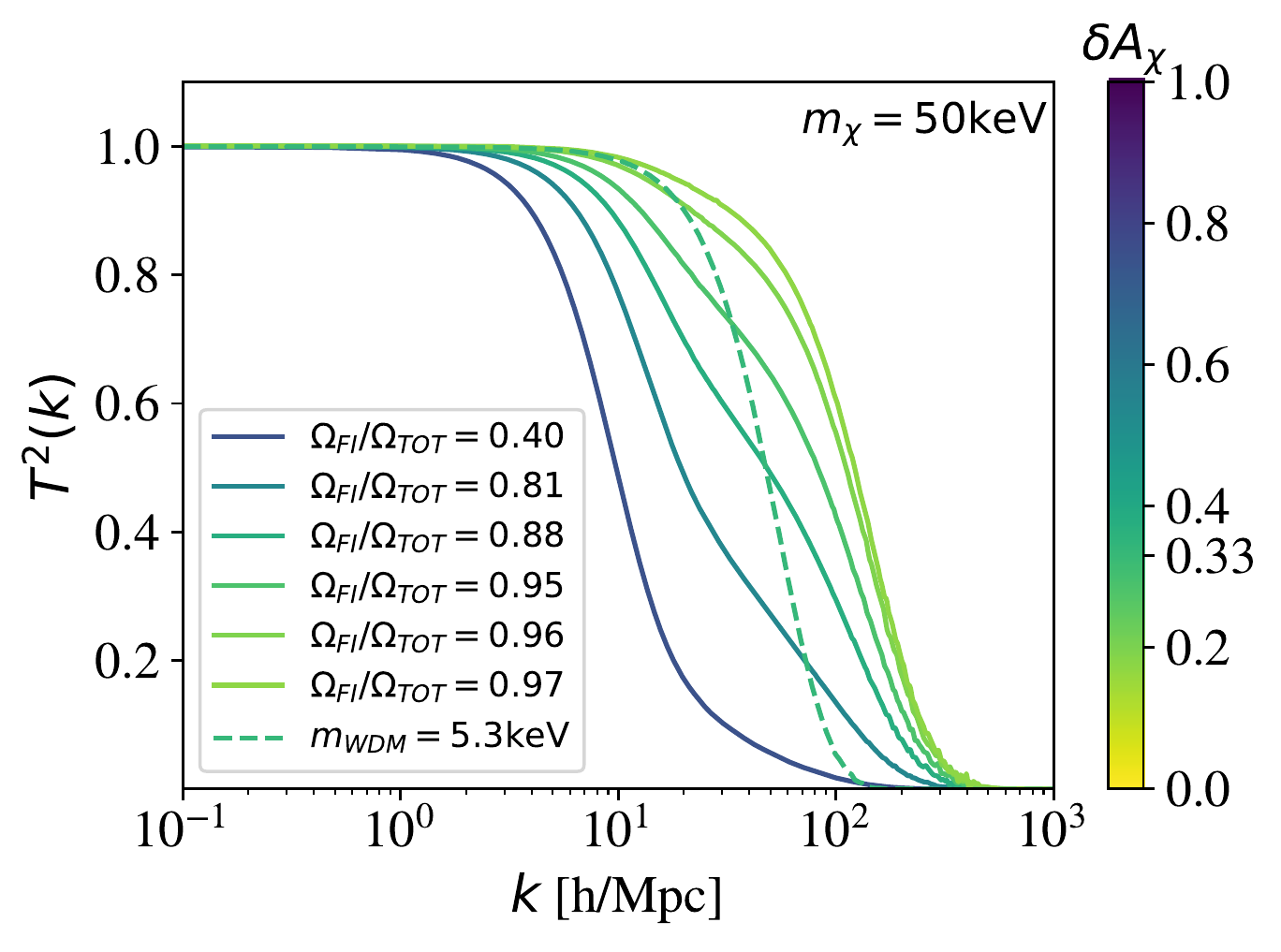}
  \vspace*{-5mm}
  \end{center}
\caption{ \textbf{Left}: Distribution functions, $q^2 g_{\chi}f_\chi(q)$, as a
  function of the rescaled momentum~$q$ for an example FIMP model.
  \textbf{Right}: Corresponding transfer functions (continuous
  coloured curves) as a function of the wavenumber $k$. The thermal
  WDM transfer function for $m_{\rm WDM}^{\mathrm{Ly}\alpha} = 5.3$
  keV (dashed curve) is also shown for comparison. These curves are
  obtained assuming the DM model considered in
  Sec.~\ref{sec:topphilic} and choosing the DM mass to be $50$
  keV. Fixing $\Omega_\chi h^2=0.12$, the remaining model parameters
  were varied such that the relative FI contribution ranges from 40\%
  to 97\%, see Sec.~\ref{sec:topphilic} for details. Going from purple
  to green indicates going from warmer to colder DM or equivalently
  going to a smaller value of the area criterion parameter $\delta A_\chi$,
  see Sec.~\ref{sec:mixed} for details.}
\label{fig:transfers}
\end{figure*}

In the case of FIMPs from the FI and SW production mechanisms, the
transfer function can take multiple forms. In the right panel of
Fig.~\ref{fig:transfers}, we illustrate the transfer functions
computed with \class~from the distributions shown in the left
panel. They correspond to different benchmark scenarios all giving
rise to $\Omega_\chi h^2=0.12$ within the top-philic DM model
described in Sec.~\ref{sec:topphilic}.  Each benchmark has a
different FI relative contribution to the total DM relic abundance
ranging from 40\% (dark purple) to 97\% (light green). This is
visible in the left panel as the FI contribution to the DM
distribution function, that peaks around $q=2.5$, increases in
amplitude when going from the dark purple to the light green
curve. In the right panel, we see that when \eg~the FI contribution
tends to 100\%, the FIMP transfer function recovers the shape of a
thermal WDM transfer function, depicted with a dashed curve for
$m^{\mathrm{Ly}\alpha}_{\rm WDM}=5.3$ keV. This has already been pointed out
in earlier works, see~\cite{Boulebnane:2017fxw,Ballesteros:2020adh}.
Similarly, for 100\% SW contribution, the FIMP transfer function
resembles a WDM-like shape. For intermediate relative FI or SW
contribution though, the shape of the transfer function can strongly
deviate from thermal WDM like scenarios.

Using our modified \textsc{class}~version, we have checked that the
transfer function of eq.~(\ref{eq:twdm}) provides a very good fit to
the case of DM produced purely through the FI or SW mechanisms. For the
fitting curves using $\mu=1.12$, as in the thermal WDM case, we
obtain
\begin{eqnarray}
  	\alpha_{\rm{FI,\,dec}} 
	&=&  0.164\q \frac{m_\chi}{1\,\text{keV} }\times \frac{1}{\delta}\w^{-0.833} \left(\frac{\gtod}{\gs(\TFI)}\right)^{0.278} h^{-1}\text{Mpc}\,,	\label{eq:alphFI}\\
        \alpha_{\rm{SW}} 
	&=&  0.0542  \q \frac{m_\chi}{1\,\text{keV}}\times\frac{\sqrt{R_\Gamma^{\rm SW}}}{ \delta } \w^{-0.833} \left(\frac{\gtod}{\gs(\TSW)}\right)^{0.278} h^{-1}\text{Mpc}\,,	\label{eq:alphSW}
\end{eqnarray}
where the parameter dependency of the breaking scales was inspired by
the analytic estimate of the Lyman-$\alpha$ bound of
eq.~(\ref{eq:Kamada}). The numerical prefactors in
eqs.~(\ref{eq:alphFI}) and~(\ref{eq:alphSW}), on the other hand, have
been obtained by doing a one-parameter fit based on the actual
transfer functions produced by \class. In the case of FI, the fit was
done over 15 models, with a final error on the prefactors of $\sim
1.5\%$. In the case of SW, we used 20 models for the fit, with an
expected error of $\sim 2\%$. In both cases, the fit has been
optimised in the mass range where we expect the Lyman-$\alpha$
constraints to appear, based on eq.~(\ref{eq:limsly}) (see
App.~\ref{sec:fit_fluid} for further discussions).

In similar spirit to what was done in \eg~\cite{Baldes:2020nuv},
we can now compare the breaking scales ($\alpha_\mathrm{FI}$ and
$\alpha_\mathrm{SW}$) found in eqs.~(\ref{eq:alphFI})
and~(\ref{eq:alphSW}) with the breaking scale for WDM
($\alpha_\mathrm{WDM}$) from eq.~(\ref{eq: alphaWDM}), to obtain
approximate Lyman-$\alpha$ bounds on FIMP DM. Assuming once more that
the DM is produced at 100\% through the FI or SW mechanisms and that this
accounts for all of the DM abundance, taking the stringent WDM limit
$m_{\rm WDM}^{\mathrm{Ly}\alpha}=5.3$ keV
from~\cite{Palanque-Delabrouille:2019iyz}  we get $m_{\rm
    FI}^{\rm lim}= 15$ keV and $m_{\rm SW}^{\rm lim}= 3.9$ keV in
  eq~(\ref{eq:limsly}), see the second line of
  Tab.~\ref{tab:NCDMnounds}.  We can see that these bounds are in very
  good agreement with the approximate constraints found in
  Sec.~\ref{sec:approx-lyman-al} using the rms velocity.  When
  using the conservative bound of $m_{\rm WDM}^{\mathrm{Ly}\alpha}=1.9$ keV
  from \cite{Garzilli:2019qki} these prefactors reduce to $m_{\rm
    FI}^{\rm lim}=3.5$ keV and $m_{\rm SW}^{\rm lim}<1.0$ keV.

\subsubsection{Area criterion}
\label{sec:mixed}

An alternative approach to extract the Lyman-$\alpha$ bounds on NCDM
scenarios is based on the area criterion introduced in~\cite{Murgia:2017lwo}, see
also~\cite{Schneider:2016uqi,DEramo:2020gpr,Egana-Ugrinovic:2021gnu}. The
methodology goes as follows. For a given DM scenario $X$, the 3D power
spectrum $P_X(k)$ has to be computed.  The deviation from the
corresponding CDM scenario is obtained by evaluating the ratio
\begin{equation}
r(k)= \frac{P_{1\mathrm{D}}^{X}(k)}{P_{1\mathrm{D}}^{\rm CDM}(k)}\quad {\rm with} \quad P^X_{1\mathrm{D}}(k)=\int_k^{\infty} \D k' \, k' \, P_X(k')\,,
  \label{eq:P1D}
\end{equation}
where $P^X_{1\mathrm{D}}$ is the 1D power spectrum in the DM scenario
$X$.

This ratio is estimated over the range of scales probed by the
Lyman-$\alpha$ observations. In~\cite{Murgia:2017lwo} the suggested
range corresponding to the MIKE/HIRES+XQ-100 combined data set, used
in~\cite{Irsic:2017ixq} to derive the stringent WDM bound considered
here, was taken to be
\begin{equation}
  [k_{\rm min},k_{\rm max}]=[0.5 \,\mathrm{h}/{\rm Mpc},20 \, \mathrm{h}/{\rm Mpc}]\,.
\label{eq:Ly-range}
\end{equation}
More precisely, in order to quantify the suppression of the power
spectrum in the NCDM model $X$, one should compute the area estimator
\begin{equation}
  \delta A_X=\frac{A_{\rm CDM}-A_X}{A_{\rm CDM}} \quad {\rm with} \quad A_X=\int_{\rm k_{\rm min}}^{k_{\rm max}} \D k' \, r(k')\,,
\end{equation}
and $A_{\rm CDM}= k_{\rm max}-k_{\rm min}$ by definition.

As underlined by the authors of the original
work~\cite{Murgia:2017lwo} introducing this criterion, let us
emphasise that the area criterion has some arbitrariness in defining
the integration limits, and should, therefore, only be used after
careful calibration with an example WDM model.  For the cosmological
and precision parameters considered in our analysis, we get
\begin{equation}
  \delta A_{\rm WDM}=0.33\quad {\rm for} \quad m_{\rm WDM}=5.3\, {\rm keV}.
  \label{eq:dAw}
\end{equation}
A NCDM scenario that would give rise to $\delta A_X=\delta A_{\rm
  WDM}$ above is thus expected to saturate the stringent WDM
Lyman-$\alpha$ bound considered here.\footnote{Notice that
  in~\cite{Murgia:2017lwo}, a much smaller $\delta A_{\rm WDM}$ of
  0.21 is reported for a 5.3 keV WDM\@. We have checked together with
  R. Murgia of~\cite{Murgia:2017lwo} that the methodology followed
  here is perfectly correct.  A  discrepancy with the numerical
  results for $\dAw$ quoted in~\cite{Murgia:2017lwo} has also been
  reported in \eg~\cite{DEramo:2020gpr}. This emphasises the
importance of recomputing self-consistently the $\dAw$ before applying
any constraint to a new NCDM scenario.}

Making use of the linear 3D power spectrum computed with our modified
version of \class~ for pure FI and SW DM scenarios and,
comparing $\delta A_\mathrm{FI,\,dec}$ and $\delta A_{\rm SW}$ to the
stringent bound provided by eq.~(\ref{eq:dAw}), we get a limit similar
to the one derived in Secs.~\ref{sec:approx-lyman-al}
and~\ref{sec:pure-fi-sw}. More precisely, for the prefactors of
eq.~(\ref{eq:limsly}), we get $m_{\rm FI}^{\rm lim}= 15$
keV\footnote{Note that if we make use of the perfect fluid approximation in
 \class, we obtain $m_{\rm FI}^{\rm lim}= 16$ keV, as 
 in~\cite{DEramo:2020gpr}. However, we will switch this approximation off
 for NCDM from FI, see App.~\ref{sec:fit_fluid}.}
and $m_{\rm SW}^{\rm lim}= 3.8$ keV, see the third line of
Tab.~\ref{tab:NCDMnounds}. We have also checked that using the fits
provided in eqs.~(\ref{eq:alphFI}) and (\ref{eq:alphSW}) instead of
the $P(k)$ from \class~gives rise to the same conclusions.  It
appears, therefore, that in the case of pure FI or SW, all 3
methodologies considered in Sec.~\ref{sec:lyman-alpha} agree with each
other. In particular, this suggests that a very accurate estimate of
the Lyman-$\alpha$ bound, for FIMP scenarios with unimodal
distribution functions, can readily be extracted from
eq.~(\ref{eq:Kamada}) without going through the detailed
implementation of the NCDM model in \class. In contrast, a more
advanced approach proposed in~\cite{Archidiacono:2019wdp}, where
a Lyman-$\alpha$ likelihood was developed for multiple NCDM models,
allows for full Monte Carlo Markov Chain (MCMC) analyses. However,
such analyses are hindered by the execution speed of the corresponding
NCDM model in \class.  For the models considered here, the
corresponding runtime needed to calculate the matter power spectrum is
of the order of $\sim 30\,$min per model\footnote{See
  App.~\ref{sec:fit_fluid} for more details on the computation time.},
making MCMC analyses computationally infeasible.  As such, here we
limit ourselves to the more simplistic methods discussed above.

In the case of mixed FI-SW scenarios, the DM distribution function is
multimodal and the resulting transfer function can significantly
deviate from the WDM one, as illustrated in
Fig.~\ref{fig:transfers}. The area criterion is the only estimator of
the Lyman-$\alpha$ bound that has been carefully tested against
hydrodynamical simulations for a large ensemble of NCDM scenarios,
see~\cite{Murgia:2017lwo,Murgia:2019igt}. For this reason, we make use
of the latter criterion when considering mixed FI-SW models. In
particular, for the set of benchmarks of Fig.~\ref{fig:transfers}, the
gradient of colours in the curves corresponds to a value of the area
criterion.  More precisely, going from purple to green curves we have
$\delta A_{\chi}= \q 0.75,0.54,0.38,0.29,0.20,0.18\w $, respectively,
\ie~the first three benchmarks are excluded when considering
eq.~(\ref{eq:dAw}).  We have also checked that the area criterion 
gives rise to a more conservative bound than the
estimator of eq.~(\ref{eq:Kamada-mix}) for mixed scenarios.  As a
result, for mixed FI-SW scenarios, it necessary to implement the exact
NCDM model in \class~in order to extract a reliable estimate of the
Lyman-$\alpha$ bound.

\subsection{Bound from $\Delta \Neff $}
\label{sec:delta-neff}

The FIMPs considered here can potentially affect the effective number
of relativistic non-photonic species, $\Neff$, entering in the computation of CMB and BBN  observables, see \eg~\cite{Merle:2015oja,
  Baumholzer:2019twf,Ballesteros:2020adh}. Here we
consider the possibility for the DM candidates to contribute
as an extra fermionic species.  Our goal is, therefore, to compute their
$\Delta \Neff (T)$ contribution at a given temperature $T$,
corresponding to a given scale factor $a(T)$. It is instructive to
first estimate for which mass range FIMPs arising from FI or SW are
still relativistic. This is the case when the rescaled momentum $
\langle q_\star\rangle $ is larger than the ratio $m_\chi/T_\star$.
Using eqs.~(\ref{eq:TNCDM}) and (\ref{eq:qstarav}), the condition on
the FIMP mass becomes
\begin{equation}
m_\chi>
   \begin{dcases*}
   \frac{\delta}{a\q T\w} \times 2\times 10^{-7}\, {\rm keV}  & for relativistic FIMP from  FI,\\
   \frac{\delta}{a\q T\w R_\Gamma^{1/2}}  \times 5\times 10^{-8}\, {\rm keV} & for relativistic FIMP from  SW,\,
   \end{dcases*}
   \label{eq:relativ}
\end{equation}
when $g_{*S} \q T_{\rm prod}\w = 106.75$. From
Sec.~\ref{sec:lyman-alpha}, we know that for FIMPs from FI,
Lyman-$\alpha$ forest data imply a lower bound on their mass of around $15$
keV\@. FIMPs from FI with larger masses cannot be further constrained
by $\Delta \Neff$ bounds from CMB data, as they are expected to
be highly non-relativistic for $a(T_{\rm CMB})\sim 10^3$. For FIMPs
from SW, with mass $m_\chi> 10$ keV we would need $R_\Gamma> 10^{-14}$
for them to be non-relativistic at CMB time. On the other hand, since
$a\q T_{\rm BBN}\w\sim 10^{-10}$, one can more easily get relativistic
FIMPs from both FI and SW at BBN time. We will, therefore, focus on
$\Delta N_{\rm eff}$ at BBN time and impose the bound~\cite{Hufnagel:2017dgo} 
\begin{equation}
  \Delta N_{\rm eff}(T_{\rm BBN})< 0.31\,,
  \label{eq:NeffBBN}
\end{equation}
at 95\% CL, see also~\cite{Pitrou:2018cgg,Patrignani:2241948}.

We compute the FIMP contribution to the effective number of
relativistic non-photonic species, $\Delta \Neff \q T\w$,
following~\cite{Merle:2015oja}.  In general, at a given bath
temperature $T$, we should evaluate
\begin{eqnarray}
  \Delta \Neff (T)&=&\frac{\rho_\chi (T)- m_\chi n_\chi(T)}{\rho_{\mathrm{rel} \, \nu} (T)/N_{\mathrm{eff}}^{\nu}}\cr
    &=&g_\chi\frac{60}{7 \pi^4}  \left(\frac{T_\star}{T_\nu}\right)^{4} \times \int \D q_\star q_\star^2 \left(\left(q_\star^2+\frac{m_\chi^2 }{T_\star^2}\right)^{1/2}-\frac{m_\chi}{T_\star}\right) f(q_\star)\,,
\label{eq:Neff}
\end{eqnarray}
where $\rho_{\mathrm{rel} \, \nu} /\Neff^{\nu}= 2\times \frac{7}{8}
\frac{\pi^2}{30} T_\nu(T)^4$ is the energy density per relativistic standard model
neutrino and $T_\star, T_\nu$ are time-dependent variables. For
relativistic FIMPs at BBN time, this contribution reduces to
\begin{eqnarray}
  \Delta \Neff^{\mathrm{rel}} \q T_{\rm BBN}\w
  &\simeq&\left.\frac{\rho^{\mathrm{rel}}_\chi }{\rho_{\mathrm{rel} \, \nu} /N_{\mathrm{eff}}^{\nu}}\right|_{ T_{\rm BBN}} =\langle q_\star\rangle \times\left.\frac{ T_\star n_\chi }{\rho_{\mathrm{rel} \, \nu}/N_{\mathrm{eff}}^{\nu}}\right|_{ T_{\rm BBN}}\,.
 \label{eq:Neffrel}
\end{eqnarray}
Rescaling $T_\star n_\chi$ from BBN time to today, keeping in mind
that $\langle q_\star\rangle$ is constant, and that $T_\nu\q T\w = T$ at BBN time, we get
\begin{equation}
  \Delta \Neff^{\mathrm{rel}}\q T_{\rm BBN}\w\simeq 5.0\times10^{-4} \times \langle q_\star\rangle {\tt T_{ncdm}} \times \left(\frac{\Omega_\chi h^2}{0.12}\right) \left(\frac{10 \, \rm keV}{m_\chi}\right)\,,
\end{equation}
where $\Omega_\chi h^2$ is the non-relativistic FIMP abundance today.
Using the values of $\langle q_\star\rangle {\tt T_{ncdm}}$ obtained
above and the condition eq.~(\ref{eq:NeffBBN}) we find
\begin{equation}
m_\chi \gtrsim 
   \begin{dcases*}
 1.3 \times 10^{-2}\, {\rm keV} \times \delta   \left(\frac{\Omega_\chi^{\rm FI} h^2}{0.12}\right)  \left( \frac{106.75}{\gs(\TFI)} \right)^{1/3} & for FI ,\\
3.4 \times 10^{-3}\, {\rm keV}\times \delta \left(R_\Gamma^{\rm SW}\right)^{-1/2} \left(\frac{\Omega_\chi^{\rm SW} h^2}{0.12}\right)   \left( \frac{106.75}{\gs(\TSW)} \right)^{1/3}& for SW,\\
   \end{dcases*}
   \label{eq:NeffFIMP}
\end{equation}
where $\Omega_\chi^{\rm FI, SW}$ refers to the FIMP abundance arising
from FI or SW production.\footnote{ The FI constraint in
  eq.~(\ref{eq:NeffFIMP}) can be applied to FI through scatterings by
  setting $\delta = 1$ when the conditions to extract
  eq.~(\ref{eq:qstarav}) are met.} At first sight, the above
constraints seem less constraining than Lyman-$\alpha$, in
agreement with~\cite{Ballesteros:2020adh}, see also~\cite{Li:2021okx}. Notice, however, that
contrarily to \eg~eq.~(\ref{eq:limsly}), the above constraints are
applicable even when ${\Omega_\chi h^2}<{0.12}$.\footnote{Also notice
  that our results for SW do not agree with the results
  of~\cite{Baumholzer:2019twf}. Let us re-emphasise, however,
  that~\cite{Baumholzer:2019twf} assumed that $B$ is kinetically
  decoupled after FO, which is usually not the case if $B$ is charged
  under standard model symmetries.  }

\section{FIMPs within a top-philic mediator model}
\label{sec:topphilic}

For an application of the above results and their comparison to other constraints we
consider a simplified $t$-channel mediator DM model. It supplements the standard model with a singlet Majorana fermion, $\chi$, and a coloured scalar mediator, $\tilde t$, with gauge quantum numbers identical to the right-handed top quark. Imposing a $Z_2$ symmetry under which $\chi\to-\chi$ and $\tilde t\to -\tilde t$ (while standard model particles transform evenly), $\chi$ is stable for $m_\chi<m_{\tilde t}$ and, hence, constitutes a viable DM candidate. 
The renormalisable interactions allowed by the $Z_2$ and gauge symmetries are described by the Lagrangian
\begin{equation}
    \mathcal{L}_\text{int} = |D_\mu \tilde t|^2 + \lambda_\chi \tilde t\, \bar{t}\,\frac{1-\gamma_5}{2}\chi +\text{h.c.} +\lambda_{H \tilde t} \,\tilde t^\dag\tilde t H^\dag H \,,
    \label{eq:stopmodel}
\end{equation}
where $D_\mu$ is the covariant derivative, $t$ the top quark Dirac field and $H$ the standard model Higgs doublet. 
The masses $m_\chi,m_{\tilde t}$ and the coupling $\lambda_\chi$ are the phenomenologically relevant parameters considered here. The latter governs the (feeble) DM interactions with the thermal bath. The Higgs portal coupling, $\lambda_{H \tilde t}$, affects the interactions of the mediator with the thermal bath. For DM production via FI, during which the mediator is in thermal equilibrium, the presence of this coupling does not affect the relevant dynamics. For the case of SW production, it can contribute to the mediator annihilation during its freeze-out, potentially lowering its abundance. However, to compete with the annihilation rate associated with the strong interactions of the mediator (which are further enhanced through non-perturbative effects, see below) requires the Higgs portal coupling to be very large. Here we assume $\lambda_{H \tilde t}$ to be well below unity, in which case it is totally negligible for the phenomenology considered.

The model is reminiscent of a supersymmetric standard model. In fact, it may be realised as a limiting case of a non-minimal supersymmetric extension in which $\chi$ is a mixture of the bino and the fermionic component of an additional supermultiplet that is a singlet under the standard model gauge group~\cite{Belanger:2005kh,Ibarra:2008kn}. However, we will remain agnostic to a possible theoretical embedding of the simplified model, assuming that the above Lagrangian captures the relevant physics. 

In the context of FI and SW production, this model has been studied in~\cite{Garny:2018ali}. Similar results have been obtained for other spin-assignments~\cite{Belanger:2018sti,Calibbi:2021fld}. A variant of the model without an imposed $Z_2$ symmetry was discussed in~\cite{Arcadi:2013aba,Arcadi:2014tsa}, while its phenomenology in the case of thermalised DM can be found in~\cite{Ibarra:2015nca,Garny:2018icg}.

In~\cite{Garny:2018ali}, constraints from Lyman-$\alpha$ forest observations have been estimated with a comparison of the respective limits on the free-streaming length obtained for WDM from~\cite{Baur:2017stq}. Here we revisit the phenomenology and improve the analysis with respect to~\cite{Garny:2018ali} in two main aspects. First, we improve the structure formation bounds utilising the methodology outlined in Secs.~\ref{sec:FIbasics} and~\ref{sec:cosmo}. In particular, computing the DM phase-space distribution and making use of the area criterion, we can derive a reliable and more stringent bound in the region of mixed FI and SW production.  Second, we take into account bound state formation effects in the mediator freeze-out, which are relevant for the SW production of DM.

In the following we will first discuss the mediator freeze-out in Sec.~\ref{sec:bound-state}. We will then detail the FI and SW production processes of DM within the model in Sec.~\ref{sec:DMprod}, before deriving the constraints on the model parameter space in Sec.~\ref{sec:pamaspace}.

\subsection{Mediator freeze-out}
\label{sec:bound-state}

For parameter regions with a sizeable SW contribution to the DM production, the DM density (and, in general, its phase-space distribution) depends on the evolution of the mediator abundance governed by thermal freeze-out. This process is subject to non-perturbative effects. On the one hand, gluon exchange between the initial state mediators modifies their wave function, leading to an enhancement with respect to the tree-level annihilation rate at small relative velocities, \ie~the Sommerfeld enhancement~\cite{Sommerfeld:1931,Hisano:2002fk,Cirelli:2007xd}. On the other hand, mediator pairs can form bound states that affect the freeze-out dynamics leading to a further reduction of the mediator abundance, see \eg~\cite{Liew:2016hqo,Mitridate:2017izz,Biondini:2018pwp,Harz:2018csl}.

Here we consider both effects in the non-relativistic limit using the computations derived in~\cite{Harz:2018csl}. Accordingly, for mediator pair-annihilation into gluons, we employ the $s$-wave annihilation Sommerfeld factor for a Coulomb potential. Mediator pair annihilation into quark pairs is $p$-wave suppressed and, hence, sub-dominant for small relative velocities. We take into account bound state formation (ionization) via one-gluon emission (absorption) and the leading bound state decay process into a pair of gluons. We consider the ground state configuration only.  
Furthermore, it is assumed that the rate of bound state number changing processes (formation, ionization or decay) is large compared to all other rates involved in the mediator freeze-out. In this case, the effects of bound state formation can be described by an effective annihilation cross-section~\cite{Mitridate:2017izz}. It reads
\begin{equation}
\label{eq:effsigmav}
\langle\sigma_{\tilde t\tilde t^\dagger}v\rangle_\text{eff} = 
		\langle\sigma_{\tilde t\tilde t^\dagger\to gg}v\rangle \times S_\text{Som}
		+ \langle\sigma_{\tilde t\tilde t^\dagger\to q \bar q}v\rangle
+\langle\sigma_{\tilde t\tilde t^\dagger\to {\cal B}g} v\rangle \times \frac{\Gamma_{\!{\cal B},\text{dec}}}{\Gamma_{\!{\cal B},\text{ion}} + \Gamma_{\!{\cal B},\text{dec}}}
\end{equation}
where $S_\text{Som}$ is the Sommerfeld enhancement factor, $\langle\sigma_{\tilde t\tilde t^\dagger\to {\cal B}g} v\rangle$ is the thermally averaged bound state formation cross-section, $\Gamma_{\!{\cal B},\text{ion}}$ is the respective ionization rate, ${\cal B} g \to \tilde t\tilde t^\dagger$, and $\Gamma_{\!{\cal B},\text{dec}}$ its decay rate, ${\cal B}  \to gg$. For further details see App.~\ref{sec:somBSF}. We compute the thermally averaged annihilation cross-sections for the perturbative processes $\tilde t\tilde t^\dagger\to gg,  q \bar q$ with \textsc{MadDM}~\cite{Ambrogi:2018jqj}. 

\begin{figure*}[t]
  \begin{center}
    \includegraphics[width=.49\textwidth,trim= {0.3cm 0.2cm 0.5cm 0.5cm},clip]{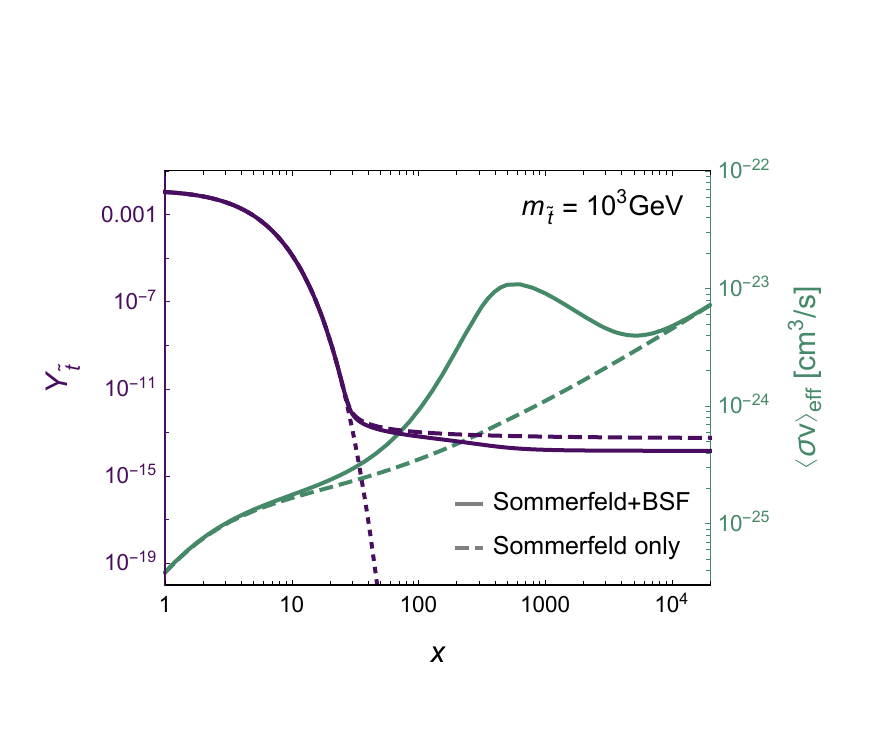}
    \includegraphics[width=.49\textwidth,trim= {0.3cm 0.2cm 0.5cm 0.5cm},clip]{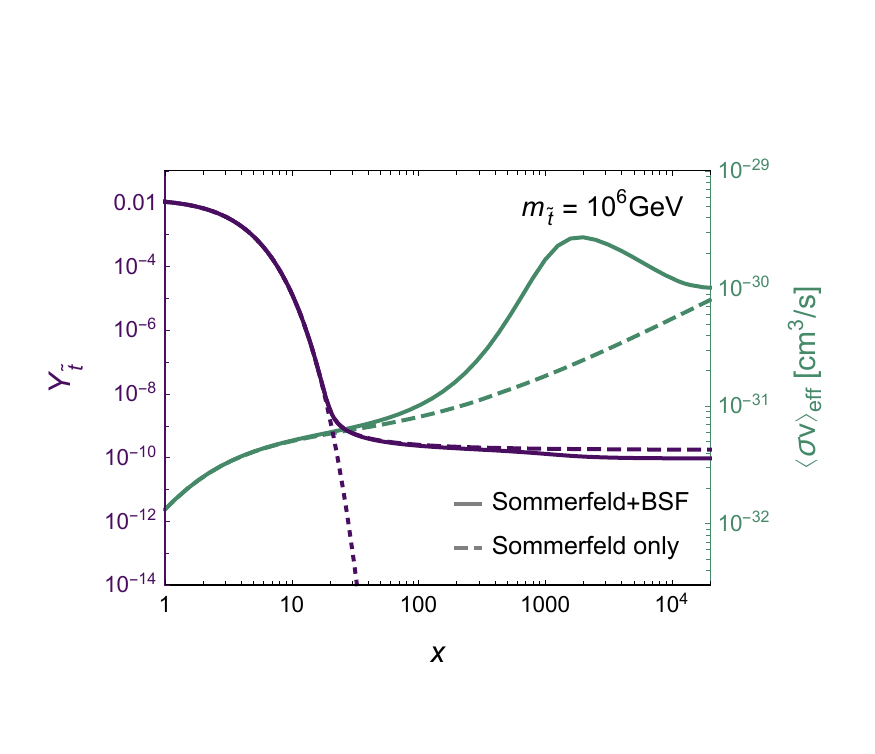}
  \vspace*{-10mm}
  \end{center}
\caption{
Evolution of the mediator abundance $Y_{\tilde t}$ (purple curves, left axes) and the effective annihilation cross-section $\langle\sigma_{\tilde t\tilde t^\dagger}v\rangle_\text{eff}$ (green curves, right axes) as a function of $x=m_{\tilde t}/T$ for $m_{\tilde t}=10^3$\,GeV (left panel) and $m_{\tilde t}=10^6$\,GeV (right panel). The solid curves take into account Sommerfeld enhancement and bound state formation effects (`Sommerfeld+BSF') while for the dashed curves only the former has been considered (`Sommerfeld only'). The purple dotted curves denote the mediator equilibrium abundance $Y_{\tilde t}^\eq$. }
\label{fig:Ymed}
\end{figure*}

Assuming the maintenance of kinetic equilibrium via elastic gluon
scattering, $\tilde t g\to \tilde t g$, throughout the entire
freeze-out process, we compute the mediator abundance by solving the
integrated Boltzmann equation\footnote{When the annihilations become
  inefficient, only the second term in the squared parenthesis of eq.~(\ref{eq:YtildetBME}) is left
  and we recover eq.~(\ref{eq:YB}) taken into account for the SW mechanism with $B=\tilde t$ in the $x>1$ limit.}
\begin{equation}
\label{eq:YtildetBME}
\frac{\mbox{d} Y_{\tilde t}}{\mbox{d} x}=\frac{1}{ 3 H}\frac{\mbox{d} s}{\mbox{d} x}\,\left[\frac12
		\langle\sigma_{\tilde t\tilde t^\dagger}v\rangle_\text{eff}\left(Y_{\tilde t}^2-{Y_{\tilde t}^\eq}^{2}\right) +  \frac{ \Gamma_{\tilde t}}{s} \,Y_{\tilde t}\right]\,,
\end{equation}
where $Y_{\tilde t}$ denotes the summed abundance of the mediator and its antiparticle and $ \Gamma_{\tilde t}$ is the (thermally averaged) rate for the mediator decay, \ie~for $ {\tilde t}\to t \chi$. Figure~\ref{fig:Ymed} shows the effective annihilation cross-section and the resulting evolution of $Y_{\tilde t}$ for two example mediator masses including Sommerfeld enhancement only (dashed curves) and including bound state effects in addition (solid curves). In these plots, we choose $ \Gamma_{\tilde t}$ small such that the decay is inefficient in the displayed $x$-range. The presence of bound states leads to a prolonged freeze-out process, as bound state effects cause an enhancement of $\langle\sigma_{\tilde t\tilde t^\dagger}v\rangle_\text{eff}$ at large $x$. Towards larger mediator masses, the maximum of this enhancement is shifted to higher $x$, while the effect on the mediator abundance becomes smaller. For a mass $10^3$ ($10^6$)\,GeV, bound state effects reduce the abundance by a factor of 3.9 (1.9). 

\subsection{Dark matter production processes}
\label{sec:DMprod}

The leading processes to DM production are scatterings of the form $X \tilde t \to X' \chi$, where $X,X'$ denote standard model particles, and mediator decays $ {\tilde t}\to t \chi$. The latter gives rise to both a FI and SW contribution. The respective vacuum decay rate reads
\begin{equation}
\Gamma_{{\tilde t}\to t \chi} = \frac{\lambda_\chi^2}{16 \pi m_{\tilde t}^3} \left(m_{\tilde t}^2 - m_\chi^2 - m_t^2\right) \lambda^{1/2}\!\left(m_{\tilde t}^2 , m_\chi^2 , m_t^2\right)\,,
\end{equation}
where $\lambda(x,y,z)=x^2+y^2+z^2 - 2(xy+ xz + yz)$. 
Pair production of DM of the form $X  X' \to \chi \chi$ is of higher order in the coupling $\lambda_\chi$ and, hence, neglected here.

\begin{figure*}[t]
  \begin{center}
    \includegraphics[width=.7\textwidth]{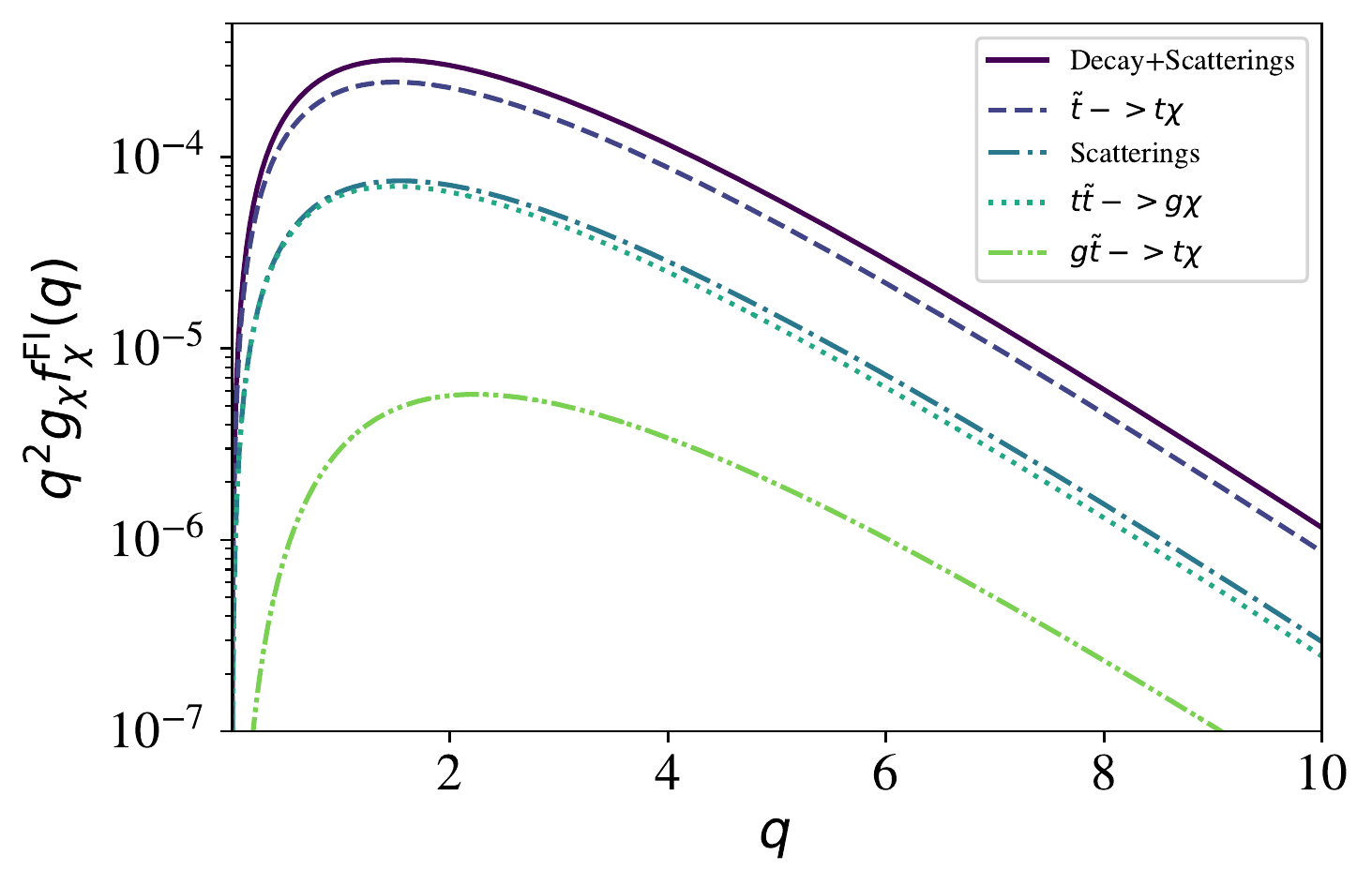}
  \vspace*{-5mm}
  \end{center}
\caption{Contributions to the DM distribution function arising from
  FI, $q^2g_\chi f_\chi^{\rm FI}(q)$, as a function of $q$ for top-philic DM
  when taking $\lambda_\chi = 10^{-7},\; m_{\tilde t} = 5.6\times
  10^6\text{ GeV},$ and $ m_\chi = 10^{-3} \text{ GeV}$. From top to
  bottom we have the total distribution arising from both decays and
  scattering (solid), as well as the decay (dashed) and the total
  scattering (dot-dashed) contributions.  The latter divides into the
  $g\tilde t \to t\chi$ (dot-dot-dashed) and the $t\tilde t \to g\chi$ (dotted)
   contributions. Because of the gluon thermal mass
  considered to regularise the scattering cross-section for $g\tilde t
  \to t\chi$, the dot-dot-dashed curve has a mean momentum shifted to
  higher values than the expected $\qav=2.5$ for FI.  }
\label{fig:FIdistrimodel}
\end{figure*}

Among the scattering processes, we consider the leading processes in the strong coupling $\alpha_\text{s}$, \ie~$t \tilde t  \to  g \chi $ and  $g \tilde t \to t \chi$, which are expected to contribute similarly. However, the second process is subject to a soft divergence, which we regularise by introducing a thermal mass for the gluon~\cite{Binder:2019ikc}; 
\begin{equation}
m_g\q T \w = \frac{4\pi \alpha_\text{s}}{6} T^2  \q N_c + \frac{N_f}{2} + \frac{N_s}{2}\w \, ,
\end{equation}
where $T$ is the bath temperature, and $N_c, N_f, N_s$ are the
number of colours and number of active fermions and scalars in the
thermal bath, respectively. The thermal mass enters the cross-section
and the lower integration limit, $\tilde s_\text{min}$, in
eq.~\eqref{eq:fscat}. Note that the process belongs to the ${\cal O}
(\alpha_\text{s})$ corrections to the mediator decay at finite
temperatures. Their rigorous computation can only be performed in
thermal field theory, which is beyond the scope of this
work.\footnote{See \eg~\cite{Biondini:2020ric,Jackson:2021dza} for recent advances in the treatment of thermal corrections relevant for FI.} Conservatively, we consider the size of the FI contribution from
scattering as a rough estimate for the uncertainty of the total FI
contribution to the relic density. The scattering processes contribute
between 15 and 25$\%$ for a mediator mass in the range of $10^3$ GeV
to $10^{10}$ GeV. Note that the thermal mass of the gluon introduces a
temperature dependence in the cross-section, as well as in the minimal
centre of mass energy. As a result, eqs.~(\ref{eq:qnavsc})
and~(\ref{eq:qnavsclim}) do not apply and the mean momentum shifts to a
higher value $\qav \approx 3.15$. However, this effect in the total
distribution is marginal because the channel $g \tilde t  \to t \chi$
is suppressed compared to the others when considering the gluon
thermal mass as a regulator. This is illustrated in
Fig.~\ref{fig:FIdistrimodel} for a parameter point with
$\lambda_\chi = 10^{-7},\; m_{\tilde t} = 5.6\times 10^6\text{ GeV},$
and $ m_\chi = 10^{-3} \text{ GeV}$. The total FI distribution is
shown with a solid curve, while the contributions from the two
scattering processes and the decay are shown with dot-dot-dashed,
dotted and dashed curves, respectively.

For very small DM masses,
the coupling $\lambda_\chi$ that yields the measured relic density can become large enough to render the decay efficient already 
close to the time of mediator freeze-out.\footnote{For the DM masses around the Lyman-$\alpha$ constraint, decays and scatterings are, however, at least about two orders of magnitude smaller than the Hubble rate for $x\lesssim 3$, justifying the commonly made approximations in the FI computation.} In this case, the distinction between the FI and SW production processes may be less obvious. For definiteness, we consider the contribution in the regime $x<7$ ($x>7$) to belong to the FI (SW) production. We only consider scatterings in the former while taking into account the full evolution of the mediator abundance, solving eq.~\eqref{eq:YtildetBME}, only in the latter regime. This value of $x$ has been chosen since scatterings are already completely negligible at this point. In addition, deviations from thermal equilibrium are still small even for the largest mediator masses considered here, which feature the earliest deviations from thermal equilibrium.
Note that the SW contribution from early decays is only comparable to the FI contribution for very large mediator masses, where the larger mediator freeze-out abundance overcompensates the small ratio of masses $m_\chi/m_{\tilde t}$ entering the SW contribution to the DM relic density.

\subsection{Viable parameter space and constraints}
\label{sec:pamaspace}

\begin{figure*}[t]
  \begin{center}
    \includegraphics[width=.465\textwidth]{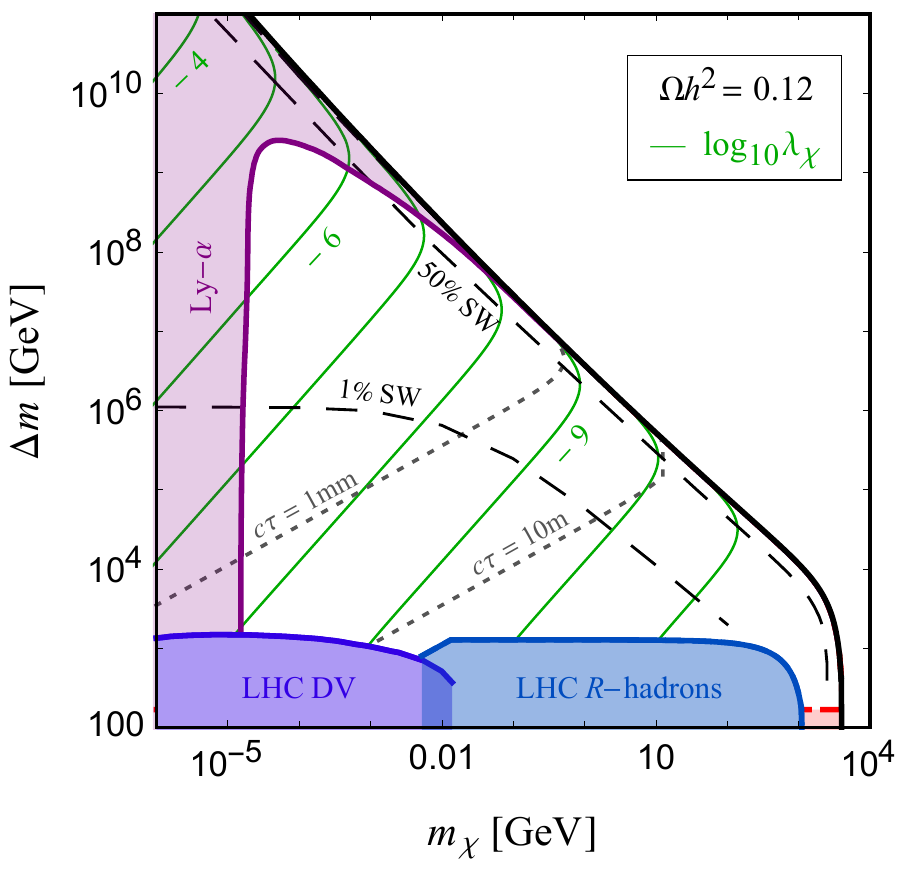}
    \hspace*{5mm}
    \includegraphics[width=.4557\textwidth]{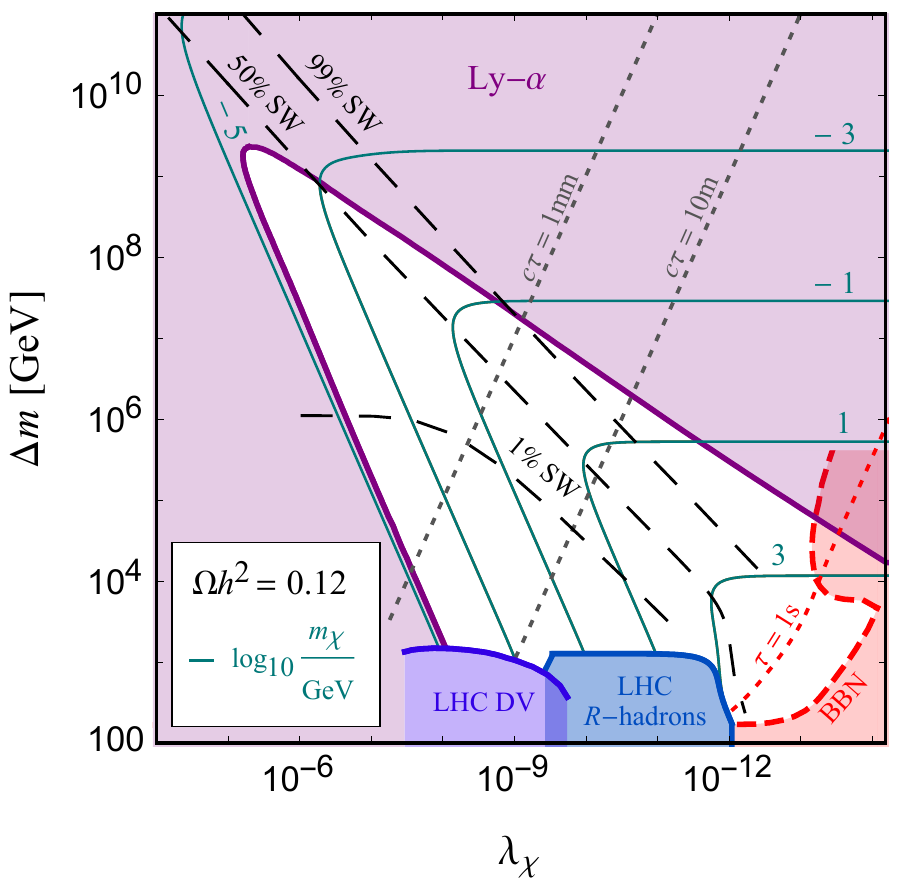}
  \vspace*{-5mm}
  \end{center}
\caption{Cosmologically viable parameter space ($\Omega h^2=0.12$) of
  the considered top-philic $t$-channel mediator model. \textbf{Left}:
  Projection onto the plane spanned by $m_\chi$ and $\Delta m=
  m_{\tilde t} - m_\chi$. The green contours denote decades of the
  coupling $\lambda_\chi$. For parameter points to the right of the
  thick black line, DM is over-abundant regardless of the coupling,
  \ie~no solution can be found. \textbf{Right}: Projection onto the
  $\Delta m$-$\lambda_\chi$-plane. The cyan contours denote decades of
  $m_\chi/$GeV. (To reduce clutter we only display every second line.)
  Note that the scale of the abscissa has been inverted allowing for a
  more direct comparison of the two projections. In both panels, the
  black, long-dashed curves denote contours of equal SW contribution
  to the total relic density. The grey dotted lines denote contours of
  equal decay length. Our constraints from the Lyman-$\alpha$
  observations (Ly-$\alpha$) are shown in purple, while BBN bounds are
  displayed in red. Constraints from LHC searches for displaced
  vertices (DV) and $R$-hadrons are shown in royal blue and aqua blue,
  respectively. }
\label{fig:paramspace}
\end{figure*}

By numerically solving 
\begin{equation}
\label{eq:lambdasol}
\Omega_\chi h^2|_\text{FI}(\lambda_\chi) + \Omega_\chi h^2|_\text{SW}(\lambda_\chi)=0.12
\end{equation}
 we compute the required DM coupling, $\lambda_\chi$, that matches the measured relic density for a given DM and mediator mass in the considered parameter space. The resulting hyperplane is shown in Fig.~\ref{fig:paramspace} by displaying contours of equal $\lambda_\chi$ in the plane spanned by $m_\chi$ and $\Delta m= m_{\tilde t} - m_\chi$ (green curves in the left panel) and by drawing contours of equal $m_\chi$ in the plane spanned by $\lambda_\chi$ and $\Delta m$ (cyan curves in the right panel). Note that we have inverted the scale of the abscissa in the right panel to make the correspondence between the two projections more obvious. To the right of the thick black line in the left panel, $\Omega_\chi h^2|_\text{SW}(\lambda_\chi)>0.12$ for any $\lambda_\chi$ and so no solution for eq.~\eqref{eq:lambdasol} can be found. Approaching this boundary from the left, the coupling drops by orders of magnitude. This region is only visually resolved in the right panel.

The black long-dashed curves denote contours of equal SW contribution. The 50\% curve divides the parameter space into the FI (to the left) and SW dominated regions (to the right). In the former the relic density is (asymptotically) proportional to $\lambda_{\chi}^2$ while in the latter the $\lambda_\chi$-dependence is mild. However, due to the prolonged freeze-out process discussed in Sec.~\ref{sec:bound-state}, even the  SW contribution depends on $\lambda_\chi$ in a considerable part of the parameter space. In particular, in the region of large mediator masses and significant SW contribution, the mediator decays while mediator pair annihilations have not yet become fully inefficient.

For the computation of the Lyman-$\alpha$ bound on the top-philic DM
parameter space, we have exploited the area criterion. As discussed in 
Sec.~\ref{sec:mixed}, this allows us to probe the mixed FI-SW scenarios 
encountered in this model. To this aim, we have used our modified version of
\class~including the analytic FI from decay and SW DM distribution
functions\footnote{Because of the prolonged freeze-out, one should a
  priori compute the DM distribution arising -- from both FI and SW --
  fully numerically by integrating out the collision term given in
  eq.~(\ref{eq:Cscat}), where $f_B$ would obtained
  using eq.~(\ref{eq:fBSW}) with $Y_B=Y_{\tilde t}$ arising from the
  integrated Boltzmann eq.~(\ref{eq:YtildetBME}). We have checked that
  using our analytic distributions of Sec.~\ref{sec:distrib} with
  $Y_{\rm FO}=\Omega_\chi h^2|_{\rm SW} (\lambda_\chi) \times
  \rho_{\rm crit}/(s_0 h^2 m_\chi)$, we recover the numerical results up
  to a few percent error.  } 
displayed in Sec.~\ref{sec:distrib},
together with fits to the numerically obtained contributions arising
from FI via scatterings. We have followed the methodology described in
Sec.~\ref{sec:mixed} for a selection of parameter points which were
expected to lie near the Lyman-$\alpha$ limits. An example of this
selection is shown in Fig.~\ref{fig:transfers} for $m_\chi=50$ keV.

The Lyman-$\alpha$ observations constrain the parameter space towards
small DM masses (in the FI dominated regime, \ie~to the left), towards
small DM couplings and, hence, large mediator lifetimes (in the SW
dominated regime, \ie~to the right), and towards large mediator masses
(in the mixed regime, \ie~to the top). The exclusion is displayed as
the purple shaded region in both panels of
Fig.~\ref{fig:paramspace}.  In the limit of FI and SW dominated
production, the limits correspond the ones in eq.~\eqref{eq:limsly} from
the area criterion. Note that the limits are considerably stronger
than the ones estimated in~\cite{Garny:2018ali}, in particular, in the
region of similar contributions from FI and SW region providing an
upper bound on the mediator of around $\Delta m = 2\times 10^9\,$GeV.

Towards small mediator masses and towards large mediator lifetimes, the 
parameter space is constrained by two further observations. First, searches 
for long-lived coloured particles at the LHC constrain mediator masses up to 
the TeV scale. Here we illustrate the limits imposed by current data 
considering searches for $R$-hadrons and displaced vertices. In the region 
of parameter space providing large lifetimes compared to the detector size, 
\ie~for $c\tau>100\,$m, we directly apply the limit from the 13\,TeV ATLAS 
search~\cite{ATLAS:2019gqq} for detector-stable $R$-hadrons containing a 
supersymmetric top-partner. For smaller lifetimes, we reinterpret the 
13\,TeV ATLAS search for displaced vertices and missing transverse 
energy~\cite{ATLAS:2017tny} within our model using the recasting 
from~\cite{Calibbi:2021fld}. We use the
squark cross-section prediction provided in~\cite{Beenakker:2016lwe} for the 
$\tilde t$ pair production at the LHC\@. Second, the decay of the coloured 
mediator during the epoch of BBN may spoil the successful
predictions for the primordial abundances of light elements~\cite{Jedamzik:2006xz,Kawasaki:2017bqm,Jedamzik:2007qk,Kusakabe:2009jt}. We estimate these 
constraints employing the results from
\cite{Jedamzik:2006xz} for a hadronic branching ratio of 1. 
The relatively mild dependence of the limits on the mediator mass is 
approximately taken into account linearly interpolating 
(and extrapolating) the results for 100\,GeV and 1\,TeV in log-log space. 
The same approach was followed in~\cite{Garny:2018ali}.

The LHC and BBN bounds are shown in Fig.~\ref{fig:paramspace} as the blue and red shaded regions, respectively. For small mediator masses, the smaller freeze-out energy density of the mediator required by eq.~\eqref{eq:lambdasol} allows for larger lifetimes. In this regime the BBN constraints arise dominantly from the observed primordial abundance of $^2$H. For larger mediator masses and correspondingly larger energy densities the stronger limits derived from $^4$He observations dominate,  constraining considerably smaller lifetimes. For comparison, we highlight the contour with a lifetime of 1s as the dotted red curve. However, as the derivation of these bounds partly rely on a extrapolation of the results of~\cite{Jedamzik:2006xz} we consider them as a rough estimate only and leave a dedicated analysis for future work. Noticeable developments of numerical tools for the reinterpretation of BBN bounds have been made more recently, see \eg~\cite{Depta:2020mhj,Arbey:2011nf}.

The LHC searches for $R$-hadrons and displaced vertices exclude
mediator masses up to around 1.3 and 1.5\,TeV, respectively.  Note
that the slight gap in their sensitivity for a DM mass between 1 and
10\,MeV -- corresponding to a mediator decay length of around 10 to
100\,m -- is expected to be closed when applying a reinterpretation of
the null-results of the ATLAS $R$-hadron search for intermediate
lifetimes. For instance, the ATLAS search in~\cite{ATLAS:2018lob}
performed for $R$-hadrons containing gluino bound states imposes
limits down to a decay length of around 3\,m that are similarly strong
as in the detector-stable regime. The null-results in the CMS search
for delayed jets~\cite{CMS:2019qjk} is expected to impose similar
constraints for intermediate lifetimes, see
\eg~\cite{Calibbi:2021fld} for a similar DM scenario.

Finally, we stress that the interplay of the above constraints is specific to
the presence of the imposed $Z_2$ symmetry that renders DM absolutely
stable. A variant of this model without a $Z_2$ symmetry has been studied, 
for instance, in~\cite{Arcadi:2013aba,Arcadi:2014tsa}. In general, allowing for
a non-zero branching fraction of the mediator decay into standard model particles only,
can lower the SW contribution to the DM density and, hence, relax the 
upper bound on the mediator mass found here. Furthermore, such a decay mode
would change the LHC bounds. However, the requirement of a sufficiently 
long DM lifetime and indirect detection limits from DM decay provide
additional constraints. A study of such scenarios is beyond the scope of this work.

\section{Conclusions}
\label{sec:concl}

Despite substantial experimental efforts dedicated to the search for
DM, no indisputable signature of DM has been found in
(astro-)particle physics experiments. As a complementary path to
unveil the nature of DM, here we explored the imprint of
non-cold DM, in the form of FIMPs, on
cosmological observables. In particular, we provided generic lower
bounds on the DM mass when DM is produced through the FI and SW
mechanisms. Our FI bound is valid for FI via 2-body decays, and we
discussed the applicability of this bound to the case of a production
via $2\to 2$ scatterings.

We first revisited the Boltzmann equations relevant for extracting the
DM momentum distribution arising from these two production mechanisms
and provided simple analytic expressions of these. Our results
are given in eqs.~(\ref{eq:fCL}).  For FI we confirmed the result
from previous literature, while the expression derived for the SW
scenario -- where DM arises from the late decay of a frozen-out
mother particle -- constitutes a new result. These analytic
expressions can also be used to describe mixed FI-SW scenarios,
where contributions from both FI and SW can be
similarly important.

Due to their relatively large velocity dispersion at the time of
structure formation, FIMPs from FI and SW production can affect
clustering on small scales. Interestingly, the associated
free-streaming effect can be constrained with Lyman-$\alpha$ forest
data, as in the case of thermal WDM\@.  For the purpose of exploiting
this probe, we implemented the analytic DM momentum distribution for
FI and SW in the Boltzmann code \textsc{class} (which we will make
publicly available). This allowed us to calculate the
linear 3D matter power spectra and the corresponding transfer
functions for both pure FI and SW DM production, as well as for mixed
FI-SW scenarios where both contributions are relevant. In the case of
pure FI and SW production, the transfer functions are similar in shape
to the one of thermal WDM\@. This enabled us to provide generic fits
to the transfer functions, the breaking scale of which depend on the DM
model parameters: the DM mass, the mother particle mass, decay
width, and the number of relativistic dof at the time of production,
see eqs.~(\ref{eq:alphFI}) and~(\ref{eq:alphSW}). These novel
results can be used to evaluate the effects of FIMP
production on the linear matter power spectrum for the pure FI and
SW scenarios, obviating the need to run a numerical Boltzmann code
such as \class.  For the mixed FI-SW scenario, however, the
corresponding distribution and transfer function can significantly
deviate from the thermal WDM case, requiring the numerical
computation.

Usually to calculate general Lyman-$\alpha$ bounds on these NCDM models, one 
should run computationally expensive hydrodynamical simulations, in order
to properly model the NCDM scenarios in the non-linear regime. Here
we instead followed three alternative approaches to estimate the
Lyman-$\alpha$ bound. The first one exploits the root mean square
velocity of the DM particles today, while the second builds on the fits to
the DM transfer functions that we provided and constrains the DM
breaking scale. The third one makes use of the area criterion, which
measures the suppression of the 1D NCDM matter power spectrum compared to 
the CDM one within the range of scales probed by the relevant cosmological
experiments. After careful calibration checks on thermal WDM, see
eqs.~(\ref{eq: alphaWDM}),~(\ref{eq:dAw}) as well as
App.~\ref{sec:fit_fluid}, we reinterpreted the existing bound from
Lyman-$\alpha$ forest observations on the WDM mass in terms of generic
lower bounds on DM mass for pure FI and SW scenarios. Our results for
each method are given in eq.~(\ref{eq:limsly}) and
Tab.~\ref{tab:NCDMnounds}, assuming a lower bound on the thermal WDM
mass given by $m_{\rm WDM}^{{\rm Ly}\alpha}=5.3$ keV\@. All three methods
are in good agreement, which can be traced back to the fact that FI and
SW production give rise to a cut in the matter power spectrum very similar
to the one of thermal WDM\@. In the case of FI from 2-body decays, we
recovered a lower bound on the DM mass of 15 keV (when $T_{\rm FI}> 
T_\mathrm{EW}$) in agreement with previous results, while the bound
from SW could exclude much larger DM masses depending on the decay width and 
mass of mother particle. For mixed FI-SW scenarios, we reached
the conclusion that the area criterion provides a conservative
estimate of the DM mass bound.

When FIMPs arising from FI and SW are still relativistic at the time
of BBN or CMB, they might provide a non negligible contribution to
$\Delta \Neff$.  We obtained a generic lower bound on the DM mass of
similar form as in the case of the Lyman-$\alpha$ bound. However,
imposing $\Delta \Neff (T_{\rm BBN})< 0.31$, the resulting bound appears 
much looser, see Tab.~\ref{tab:NCDMnounds}. Notice, though, that the latter
bound can be applied without the need of using any Boltzmann code or
hydrodynamical simulations and is also applicable in general to mixed
scenarios, see eq.~(\ref{eq:NeffFIMP}).

Having seen the general application, we turned our attention to an
example model, namely a coloured $t$-channel DM model. Here we revisited
the top-philic DM model, taking special care in the treatment of 
non-perturbative effects, such as Sommerfeld and bound state enhancement
effects on coloured mediator annihilation cross-section at early times,
as well as on the computation of the DM production via $2\to 2$
scatterings. This is of particular importance in this model in the case of 
SW and FI production, respectively. The two panels of
Fig.~\ref{fig:paramspace} summarise the viable parameter space of
FIMPs arising from FI and SW production in this scenario,
complementarily bounded by cosmological (Lyman-$\alpha$, BBN) and
particle physics (LHC $R$-hadrons and displaced vertices searches)
observables. In particular, the Lyman-$\alpha$ bound derived in the
first part of this paper plays an important role.
On the one hand, it excludes small DM masses, ${\cal O}(15$ keV), in the 
region of dominant FI production. On the other hand, 
it constrains the parameter space towards small couplings and, hence, large 
mediator lifetimes in the case of dominant SW production. In the latter 
case, Lyman-$\alpha$ observations supersede BBN constraints for mediator 
masses above $10^4$ GeV and reach DM masses up to ${\cal O}(100$ GeV).

Here we have shown the importance of structure formation bounds in
constraining FIMPs arising from FI and SW mechanisms, and illustrated
the need to consider bounds from both particle physics and cosmology
to fully understand these scenarios. In particular, the case
of mixed NCDM models -- giving rise to a multimodal momentum
distribution -- has, to our knowledge, not been discussed thoroughly in
the literature. This case can naturally appear in FIMP scenarios with a
decaying mother particle at the origin of the DM production. For the
corresponding transfer function, which significantly deviates from
the standard WDM scenario or from mixed warm + cold DM scenarios, no
example hydrodynamical simulations have been run, and we can only
provide a conservative lower bound on the DM mass. In future, it would 
be interesting to provide a thorough analysis of this case to validate our
estimations and to check if other probes, such as reionization, the
luminosity function at high redshift, or the 21 cm signal could help to test  
these models further and distinguish them from the WDM-like DM scenarios.


\acknowledgments We would like to thank S.~Junius for discussion and
providing us with his recasting of DV+MET searches.  We would also
like to thank R.~Murgia for clarifications on \mbox{Lyman-$\alpha$}
constraints for FIMPs as well as F.~D'Eramo and A.~Lenoci for
discussions.

LLH is a Research associate and QD benefits form a FRIA
PhD Grant of the Fonds de la Recherche Scientifique F.R.S.-FNRS. LLH, QD and
DH acknowledge support of the FNRS research grant number F.4520.19
and the IISN convention 4.4503.15. JH~acknowledges support from the Collaborative Research Center TRR 257 and the F.R.S.-FNRS (Charg\'e de recherches). DH is further supported by the Academy of Finland grant no. 328958.

\appendix
\section{Details of the integration of the Boltzmann equations}
\label{sec:more}

In this section, we highlight some parts of the calculations needed to
solve the Boltzmann equation given in eq.~\eqref{eq:fcoll} for
the production of FIMPs from decays. First, we provide details on
the derivation of the limits of integration in eq.~(\ref{eq:Cdec}). They arise from
the fact that the cosine of the angle between the momenta of the
decaying bath particle and the DM particle should satisfy the
condition $|\cos \theta|\leq 1$ or, equivalently,
\begin{equation}
 \Bigg|\frac{m^2_A -m^2_B - m_{\chi}^2 +2E_{\chi}E_B}{2p_Bp}\Bigg|\leq 1\\,
\end{equation}
where $p_B$ denotes the mother particle momentum and $p$ is the DM momentum.
This translates into the second order equation
\begin{equation}\label{eq:ineq}
4(p^2-E_\chi^2)E_B^2-4E_\chi \Lambda\, E_B + \Lambda^2 + 4m_B^2p^2 \leq 0 \,,
\end{equation}
with $\Lambda =m_{B}^2+m_\chi^2-m_A^2$. The two endpoints of this inequality yield the integration bounds  $\xi_{\pm}$ in eq.~\eqref{eq:Cdec}. The generic form of  these bounds, without neglecting $m_\chi$,  can be found in~\cite{Boulebnane:2017fxw}. Assuming $m_\chi\ll m_B, m_A$,
eq.~(\ref{eq:ineq}) reduces to a first order equation, yielding only
a lower bound on the rescaled energy of the bath particle in
eq.~\eqref{eq:Cdec},
\begin{equation}
\xi_B \geq  \frac{q}{\delta} + \frac{\delta x^2}{4q} = \xi_{B\, \rm min}\,.
\end{equation} 
As a result, integrating eq.~\eqref{eq:fcollx} over $x$ between some ${x_\text{min}}$ and ${x_0}$ we obtain
\begin{equation}
g_{\chi}f_{\chi}(q) = \int_{x_\text{min}}^{x_0} \D x \frac{x^2 M_0}{16\pi m_B^3 q^2}  \int_{\xi_{B\, \rm min}}^{\infty} \D \xi_B f_B|{\cal
  M}|^2_{B\to A\chi}
\label{eq:detailed}
\end{equation}
for FIMPs produced through $B$ decay.

  Below, we first further discuss  the case for DM production via the
  FI process, and then the case of SW production.

\subsection*{Freeze-in from decays}
\label{app: Decays}

In the case of FI, the FIMP is produced when the mother particle is in chemical and kinetic
equilibrium with the bath. Assuming the Maxwell-Boltzmann distribution
and setting the
lower and upper integration bounds of $x$ in
eq.~(\ref{eq:detailed}) to $0$ and $\infty$, respectively, we obtain the
analytic expression displayed in the main text, eq.~\eqref{eq:fdec}. 

Note that the lower limit of integration causes the resulting momentum
distribution in eq.~\eqref{eq:fdec} to diverge for $q\rightarrow
0$. While this divergence does not affect the physically relevant
quantities, such as the FIMP number density which involves the product
$q^2 f_\chi(q)$, we stress that the divergence is absent altogether
when the lower bound of the $x$ integration, $x_\text{min}$, is different
from zero. In this latter case, the DM momentum distribution for
production through freeze-in reads
\begin{equation}
 g_{\chi}f_{\chi}^\mathrm{FI,\,dec}(q) = \frac{2g_B R^\text{FI}_{\Gamma}}{\delta^3}\frac{\text{e}^{-q}}{q}\q x_{\rm min}\, \delta \exp\left[-\frac{x_{\rm min}^2 \, \delta}{4q}\right]+\sqrt{\pi\delta q} \;  \text{erfc}\!\left[\frac{x_{\rm min}}{2}\sqrt{\frac{\delta}{q}}\right]\w,
\label{eq:fFI-xmin}
\end{equation}
where $\text{erfc}(z)$ denotes the complementary error function.
A realistic value of $x_\text{min}$ corresponds to the reheating temperature, $T_\text{rh}$, with $x_\text{min} \sim m_B/T_\text{rh}$. The above approximation, $x_\text{min}=0$, is well justified as long as $T_\text{rh}\gg m_B$.

\subsection*{SuperWIMP case for constant $\gs$}

In Sec.~\ref{sec:SW}, for SW production, \ie~after $B$ has frozen
out and has become non-relativistic, we consider for the equilibrium
quantities 
\begin{eqnarray}
  f_B^{\eq}(x,q)&=&\exp[-\sqrt{q^2+x^2}]\label{eq:fBeq}\\
    Y_B^{\eq}(x)&=&\frac{g_B}{\gs}\frac{45}{2\pi^2}\left(\frac{x}{2\pi}\right)^{3/2} \exp[-x] \,, \label{eq:YBeq}
\end{eqnarray}
In the latter case, eq.~(\ref{eq:dfXdxR}) becomes:
\begin{eqnarray}
g_{\chi}  \partial_x  f_\chi^{\rm SW}(x,q)&=&C_\mathrm{SW}\frac{\sqrt{x} }{ q^2} \exp[x- \delta x^2/(4q)-q/\delta- R_\Gamma(x^2-x^2_\mathrm{FO})/2 ]\,.
\label{eq:dfXdxR2}
  \end{eqnarray}
Integrating over $x$, we get a DM distribution
function of the form
\begin{equation}
\begin{split}
g_{\chi} f_\chi^\mathrm{SW}=&\;g_{\chi}\int_0^\infty\text{d}x \, \partial_x f_\chi^\mathrm{SW}(x,q) \\
=&\;\frac{C_\mathrm{SW} }{\sqrt{2} q^{7/4} (\delta +2 q R_\Gamma)^{5/4}}
 \exp\!\left(\frac{R_\Gamma x_\mathrm{FO}^2}{2}-\frac{q}{\delta }\right)\times\\
 &
  \left[q \,\Gamma\!\left(\frac{1}{4}\right) {}_{\;1\!\!\;}F_1\!\left(\frac{5}{4},\frac{3}{2},\frac{q}{\delta +2 q R_\Gamma}\right)
  +2 \,\Gamma\!\left(\frac{3}{4}\right) \sqrt{q (\delta +2 q R_\Gamma)} {}_{\;1\!\!\;}F_1\!\left(\frac{3}{4},\frac{1}{2},\frac{q}{\delta +2 q R_\Gamma}\right)\right]
  \end{split}
  \label{eq:fSWfull}
\end{equation}
where ${}_{1\!\!\;}F_1(a,b,z)$ is the Kummer confluent hypergeometric function. This result can be simplified to recover the solution given in eq.~\eqref{eq:fSW} in the following way. By setting the integration bounds of $x$ to $0$ and $\infty$, we make the approximation $x_\mathrm{FO}\simeq 0$, so we can drop the $x^2_\mathrm{FO}$ term in the resulting exponential.\footnote{Similar to the case of freeze-in discussed above, setting the lower integration bound to 0 induces a formal divergence of $f_\chi (q)$ for $q\to0$ that does, however, not affect the considered physically relevant quantities. Note that the contribution in eq.~\eqref{eq:fSWfull} from small $x$ for which $Y_\text{FO}\ll Y_B^{\eq}(x)$ is totally negligible in comparison to the contribution from freeze-in. This justifies the approximations made.} In addition, we can expand the function ${}_{1\!\!\;}F_1(a,b,z)$ for large $z$, \ie~for $q \gg \delta+2 q R_\Gamma$, to be:
\begin{equation}
{}_{1\!\!\;}F_1(a,b,z)\simeq \frac{\Gamma(a)}{\Gamma(b)} z^{a-b} \exp(z)
\end{equation}
to obtain:
\begin{equation}
g_{\chi}f_\chi^\mathrm{SW}\simeq \sqrt{8 \pi } \, \frac{ C_\mathrm{SW} }{q}\, \exp\!\left(-\frac{2 q^2 R_\Gamma}{\delta (2 q R_\Gamma+\delta)}\right) \frac{1}{2   q R_\Gamma+\delta}
\end{equation}
Furthermore, assuming $2 q R_\Gamma\ll\delta$ we arrive at the simple expression of eq.~(\ref{eq:fSW}), which is the one we use in the bulk of the text in Sec.~\ref{sec:SW}.

\subsection*{SuperWIMP for varying $\gs(x)$}
\label{sec:gSx}

When the number of relativistic dof vary in time while the FIMPs are
produced, the choice of time and momentum variables in
eq.~(\ref{eq:xq-basic}) is not the most convenient. The total time
derivative of eq.~(\ref{eq:fcoll}) would indeed involve two contributions:
\begin{equation}
  \frac{\D  f_\chi}{\D t}=  x  \bar H \partial_x f_\chi+ \partial_t q \,\partial_q f_\chi\,,
  \label{eq:boltzgsx}
\end{equation}
with $\partial_t q=\partial_t \gs^{1/3} \neq 0$ and $\bar H$ was
defined in eq.~(\ref{eq:barH}). Trading $T$ with $s^{1/3}$ and
defining time and rescaled momentum variables:
\begin{equation}
x_s= \frac{m_{ref}}{s^{1/3}} \quad \mbox{and} \quad q_s= \frac{p}{s^{1/3}}\,,
  \label{eq:xsqs}
\end{equation}
we have in full generality:
\begin{equation}
  x_s  H  \partial_{x_s} f_\chi={\cal C}[f_\chi]\,,
  \label{eq:fcollxs}
\end{equation}
as $p$ and $s^{1/3}$ simply scale as $1/a$ when entropy is conserved,
see also \cite{Belanger:2018ccd,Belanger:2020npe} for similar choice
of momentum variable.  On general grounds, in eq.~(\ref{eq:fcollxs}),
we shall re-express all $q$ and $x$ variables in terms of $q_s$ and
$x_s$, we shall take into account the time-dependence of $M_0$\footnote{This allows us to rewrite $x\bar H \partial_x Y$ 
as $x_s H \partial_{x_s} Y$.} 
\begin{equation}
  \frac{\D  \ln Y_B}{\D x_s}=- x_s  R_\Gamma \frac{s^{2/3}}{T^2} \sqrt{\frac{g_*(x_s(\TSW))}{g_*(x_s)}} \frac{K_1(x_s)}{K_2(x_s)}
\end{equation}
where $R_\Gamma$ is the constant factor of eq.~(\ref{eq:Rgam}), the
ratio of $\sqrt{g_*}$ account for $M_0(x)$ dependence. We have also
explicitly written the ratio of the modified Bessel functions of the second kind, which reduces to one in the non-relativistic limit, \ie~$x_s\gg 1$.

In this paper, we consider bath particles with masses above the
TeV, \ie~with $\TFI=m_B/3>\TEW$. In this case, the constant
$\gs$ approach followed in the bulk of this paper is perfectly
correct. In contrast, the SW decay could happen much latter and end up
in a period with $\gs \ll \gs (\TEW)$. In the latter case, integrating
out numerically eq.~(\ref{eq:fcollxs}) from $x_s=x_s(\TFO)$ to
$\infty$, one ends up with a momentum distribution $ f_\chi(q_s)$
which  in turn can  be integrated out on $q_s$ to obtain the correct DM relic number density. 
The variation of the number of relativistic dof along DM production
could in particular affect the small coupling region of our viable
parameter space of Fig.~\ref{fig:paramspace} where the SW mechanism
drives the relic DM abundance. For the latter region, we have
explicitly checked that, integrating numerically eq.~(\ref{eq:fcollxs}),
no significant change in the SW distribution function is observed
compared to the analytic result derived with fixed value of the
relativistic dof in Sec.~\ref{sec:SW}.

\section{Sommerfeld enhancement and bound state effects}
\label{sec:somBSF}

In this appendix, we provide all expressions associated to Sommerfeld enhancement and bound state formation entering the computation of the effective annihilation cross-section, eq.~\eqref{eq:effsigmav}. For a derivation of these expressions and further details we refer the reader to~\cite{Harz:2018csl} and references therein.

In the Coulomb limit, the Sommerfeld enhancement factor for the $s$-wave annihilation process $\tilde t \tilde t^\dagger \to gg$ reads 
\begin{equation}
S_\text{Som} = \frac27 \,S_{0}\!\left(\frac{4\alpha_\text{s}^\text{S}}{3v_{\mathrm{rel}}}\right) + \frac57 \,S_{0}\!\left(-\frac{\alpha_\text{s}^\text{S}}{6v_{\mathrm{rel}}}\right)
\end{equation}
where
\begin{equation}
S_0 (\zeta) = \frac{2\pi \zeta}{1-e^{-2\pi \zeta}} \,.
\label{eq:S0}
\end{equation}

Employing the non-relativistic limit, the thermally averaged bound state formation cross-section and ionization rate can be expressed as
\begin{equation}
\langle\sigma_{\tilde t\tilde t^\dagger\to {\cal B}g} v\rangle = 
\left(\frac{m_{\tilde t}}{T}\right)^{3/2} \left(\frac{1}{4\pi }\right)^{\!1/2} 
\int_0^\infty \D\vrel \,\vrel^2
\, \exp \!\left(-\frac{m_{\tilde t} \vrel^2}{4T}\right) 
[1+f_g(\omega)]\,
\sigma_{\tilde t\tilde t^\dagger\to {\cal B}g} \vrel 
\label{eq:sigmavBSFav}
\end{equation}
and
\begin{equation}
\Gamma_{\!{\cal B},\text{ion}}= 
\frac{g_{\tilde t}^2 m_{\tilde t}^3}{16\pi^2 g_{\cal B}}
\int_0^\infty \D\vrel \,
\vrel^2 \,f_g(\omega)\,
\sigma_{\tilde t\tilde t^\dagger\to {\cal B}g} \vrel \,,
\label{eq:GammaIon}
\end{equation}
respectively, where $g_{\tilde t} = 3$, $g_{{\cal B}}= 1$ 
and $f_g(\omega_g) = 1/(e^{\omega_g/T} - 1)$ is the gluon occupation number, with 
\begin{equation}
\omega_g = \frac{m_{\tilde t}}{4} \left[ \left(\frac{4\alpha_\text{s}^\text{B}}{3} \right)^2 + \vrel^2\right] \,.
\label{eq:omega}
\end{equation}
We consider the ground state only. The bound state formation cross-section appearing in the above expressions reads
\begin{equation}
\sigma_{\tilde t\tilde t^\dagger\to {\cal B}g} \vrel = 
 \frac{2^7 17^2}{3^5} 
 \frac{\pi \alpha_\text{s}^\text{BSF} \alpha_\text{s}^\text{B}}{m_{\tilde t}^2}
\times S_\text{BSF}  (\zeta_\text{S}, \zeta_\text{B}) \,,
\label{eq:sigmaBSF}
\end{equation}
where 
\begin{equation}
S_\text{BSF} (\zeta_\text{S}, \zeta_\text{B}) =
S_{0}\!\left(\zeta_\text{S}\right)
 \frac{\left(1+\zeta_\text{S}^2\right) \zeta_\text{B}^4}{(1+\zeta_\text{B}^2)^3} \, \exp \left[ - 4 \, \zeta_\text{S} \, {\rm acot} (\zeta_\text{B}) \right]  ,
\label{eq:SBSF}
\end{equation}
and
\begin{equation}
\zeta_\text{S} = -\frac{\alpha_\text{s}^\text{S}}{6 \vrel}\,,\quad \zeta_\text{B} = \frac{4\alpha_\text{s}^\text{B}}{3\vrel}\,.
\end{equation}
Finally, the rate for the leading decay mode, ${\cal B}\to gg$, is
\begin{equation}
\Gamma_{\!{\cal B},\text{dec}} = \frac{32}{81}  \, m_{\tilde t} \, (\alpha_\text{s}^\text{ann})^2  (\alpha_\text{s}^\text{B})^3  \,.
\label{eq:GammaDec}  
\end{equation}
The couplings $\alpha_\text{s}^i$ in the above expressions denote the strong coupling, $\alpha_\text{s}= g_s^2/(4\pi)$, evaluated at different scales:
\begin{equation}
\alpha_\text{s}^\text{ann} = \alpha_\text{s}(m_{\tilde t})\,,\quad 
\alpha_\text{s}^\text{S} = \alpha_\text{s}\!\left(\frac{m_{\tilde t}\vrel}{2}\right)\,,\quad
\alpha_\text{s}^\text{B} = \alpha_\text{s}\!\left(\frac{4 m_{\tilde t}\alpha_\text{s}^\text{B}}{6}\right)\,,\quad
\alpha_\text{s}^\text{BSF} = \alpha_\text{s}\!\left(\omega_g\right)\,.
\end{equation}

\section{Lyman-$\alpha$ fit and fluid approximation}
\label{sec:fit_fluid}
As discussed in Sec.~\ref{sec:pure-fi-sw},~\cite{Viel:2005qj} 
obtained a very good fit for $\alpha$ and $\mu$ (introduced in 
eq.~(\ref{eq:twdm})), from dedicated N-body simulations. In the 
aforementioned reference, for a given WDM mass $m_\mathrm{WDM}$, the 
best fit is obtained for $\mu=1.12$ and 
\begin{eqnarray}
  	\alpha_{\rm{WDM}} 
		&=&  \alpha_\text{prefactor} \q \frac{m_{\rm WDM}}{1\,\text{keV}}\w^{-1.11}\q\frac{\Omega_{\rm WDM}}{0.25}\w^{0.11}\q\frac{h}{0.7}\w^{1.22}h^{-1}\text{Mpc}\,,\label{eq:viel_ori}	
\end{eqnarray}
with $\alpha_\text{prefactor} = 0.049$. While this fit performs very well at masses of $m_\mathrm{WDM} \lesssim 3$ keV, which was more than enough given the existing bounds at the time the fit was derived, at higher WDM masses the accuracy degrades, leading to an error of a few percent. By comparing the fit to the transfer functions obtained from \class, we find the following adjustments:

\begin{equation}
\alpha_\text{prefactor}  = 
   \begin{dcases*}
   0.049 \ \ \ \text{for } m_\mathrm{WDM} \lesssim 3\,\text{keV} \,, \\
   0.045 \ \ \ \text{for }  3\,\text{keV} \lesssim m_\mathrm{WDM} \lesssim 6\,\text{keV} \,, \\
   0.043 \ \ \ \text{for } m_\mathrm{WDM} \gtrsim 6\,\text{keV} \,.
   \end{dcases*}
   \label{eq:WDM_alphas}
\end{equation}

\begin{table}[t]
	\centering
	\renewcommand{\arraystretch}{1.5}
	\begin{tabular}{| l || c | c | c | c |}
		\hline
		\multirow{2}{*}{Parameter} & \multicolumn{2}{c |}{FI} & \multicolumn{2}{c |}{SW} \\ \cline{2-5}
			& Min & Max & Min & Max\\ \hline
		$m_\chi \ [\text{GeV}] $& $1.4 \times 10^{-5}$ & $1.7 \times 10^{-5}$ & $1.0 \times 10^{-1}$ & $3.2 \times 10^{2}$ \\ \hline
		$\delta$ & 0.5 & 1.0 & 0.5 & 1.0 \\ \hline
		$R_\Gamma^{\text{SW}}$ & --- & --- & $7.0 \times 10^{-17}$ & $3.6 \times 10^{-9}$ \\[1mm] \hline
	\end{tabular} 
	\caption{Parameter range for which the fits of eqs.~(\ref{eq:alphFI}) and~(\ref{eq:alphSW}) have been optimised.}
	\label{tab:ranges}
\end{table}

We note that this prescription provides a very good fit to the thermal 
WDM transfer functions obtained with \class, provided that the perfect 
fluid approximation of the code is switched off. 
As discussed in~\cite{Lesgourgues:2011rh}, \class~features a
fluid approximation for NCDM models, whereby the species is treated as 
a perfect fluid, which allows to solve the Boltzmann hierarchy quicker. 
This results in a substantial speed-up in the computation. 
However, as already discussed in~\cite{Lesgourgues:2011rh}, when 
considering smaller scales, such as those relevant for Lyman-$\alpha$ 
probes, this approximation needs to be turned off, which can be accomplished in \class~by setting \texttt{ncdm\_fluid\_approximation = 3}. The validity of this approximation was also discussed recently in the context of other NCDM models, namely DM interacting with neutrinos, in~\cite{Mosbech:2020ahp}.

As the fluid approximation needs to be turned off for improved accuracy, the computation of these models in \class~is substantially slowed down. This is further hindered by the precise $q$-sampling needed in the phase-space distribution to properly account for both FI and SW contributions in the mixed scenarios. As such, obtaining the matter power spectrum for each model takes between 20--40 minutes, which, unfortunately, makes running Markov Chain Monte Carlo simulations infeasible. This justifies our choice to find alternative methods like those described in Sec.~\ref{sec:lyman-alpha}.

Based on the fit obtained in eqs.~(\ref{eq:viel_ori}) and~(\ref{eq:WDM_alphas}), in Sec.~\ref{sec:pure-fi-sw} we derived fits for pure FI and SW models (eqs.~(\ref{eq:alphFI}) and~(\ref{eq:alphSW})). These fits have been optimised in the parameter range described in Tab.~\ref{tab:ranges}.

\bibliography{bibFIMP}{}
\bibliographystyle{JHEP}


\end{document}